\documentclass[10pt]{article}
\usepackage[utf8]{inputenc}

\usepackage{amsmath}% Advanced Math Typesetting
\usepackage{amsfonts}% Unicode support
\usepackage{amssymb}
\usepackage{makeidx}
\usepackage{graphicx}% Add pictures to document
\usepackage{caption}
\usepackage{subcaption}
\usepackage{listings}% Source code formatting and highlighting
\usepackage{apacite}
\usepackage{natbib}
\usepackage{indentfirst}
\usepackage{bbm}
\usepackage{xcolor}
\usepackage{siunitx}
\usepackage{multirow}
\usepackage{booktabs}
\usepackage{mathtools}
\usepackage{enumitem}
\usepackage{setspace}
\usepackage[margin=1in]{geometry}
 
\usepackage{authblk}
\usepackage[pdfborder={0 0 0}]{hyperref}% For email addresses
\title{Portfolio Optimization Constrained by Performance Attribution}

\author{Yuan Hu and W. Brent Lindquist}

\affil[]{%
Department of Mathematics and Statistics, Texas Tech University, Lubbock, TX 79409-1042, USA;
yuan.hu@ttu.edu, brent.lindquist@ttu.edu}

\begin{document}
%*******************************************************************************************************
\thispagestyle{plain}
\begin{spacing}{1.0}
\maketitle
\noindent {\textbf {Abstract}}
This paper investigates performance attribution measures as a basis for constraining portfolio optimization.
We employ optimizations that minimize expected tail loss and investigate both asset allocation (AA) and the selection effect (SE)
as hard constraints on asset weights.
The test portfolio consists of stocks from the Dow Jones Industrial Average index;
the benchmark is an equi-weighted portfolio of the same stocks.
Performance of the optimized portfolios is judged using comparisons of cumulative price and
the risk-measures maximum drawdown, Sharpe ratio, and Rachev ratio.
The results suggest a positive role in price and risk-measure performance for the imposition of constraints on AA and SE,
with SE constraints producing the larger performance enhancement.
\\
\\
\noindent
\textbf{Keywords} portfolio optimization; performance attribution; asset allocation; selection effect
\end{spacing}

\section{Introduction}
\label{sec1}
\noindent 
How well a portfolio performs is always the major concern for investors,
and is usually the major metric reflecting investor confidence in the portfolio's management.
In common terms, a good portfolio delivers satisfactory return with low risk.
Attribution analysis provides measures for how well an portfolio is being managed.
Paraphrasing from Bacon (\citeyear{Bacon2008}), performance attribution is a technique used to quantify the excess return
(relative to a benchmark) of a portfolio and explain that performance in terms of investment strategy and market conditions.
From a management perspective, attribution analysis has been used to monitor performance,
to identify early indications of underperformance,
and gain investor confidence by demonstrating a thorough understanding of the performance drivers.

Following the fundamental work on performance attribution by \cite{Brinson1985} and \cite{Brinson1986},
we decompose excess return into two quantities that reflect investment strategy: asset allocation (AA),
which measures the contribution of each asset class in a portfolio to total performance of the portfolio,
and the selection effect (SE), which measures the impact of choice of assets within each class in the portfolio.
As is apparent from their definitions in the next section, AA and SE measure the differences between mean performance of
asset classes in a managed portfolio and those of a market benchmark, and are therefore `blind' to volatility effects, i.e. to tail-risk.
Motivated by this and by the work of \cite{Biglova2007} and \cite{Rachev2009},
we investigate the impact on portfolio optimization using AA and SE as hard constraints on asset weights as a method
of combining performance and tail-risk control.
We apply this methodology to a test portfolio of stocks comprising a major market index;
specifically the Dow Jones Industrial Average.
Optimization is performed by minimizing expected tail loss (ETL) at a specified quantile level, $\alpha$.
For the required market benchmark  we utilize an equi-weighted portfolio comprised of the same assets.
Performance of the resulting optimal portfolios is measured in terms of cumulative portfolio price and standard risk-measures.

Section \ref{sec2} of the paper discusses performance attribution, the constrained ETL portfolio optimizations and
the risk-measures employed to gauge their performance.
Section \ref{sec3} presents the results of the methodology applied to the test portfolio.
Section \ref{sec4} summarizes our conclusions.
%***************************************************************************

\section{Methodology}
\label{sec2}
\noindent
Consider a managed portfolio $p$ comprised of $N$ assets, consisting of $M$ asset classes with $n_i$ assets in class $i;\  i = 1, \dots, M$,
such that $\sum_{i=1}^M n_i=N$.
Let $b$ denote a benchmark portfolio comprised of the same assets.
Let the index pair, $ij; \ i = 1, \dots, M; \ j=1,\dots,n_i$, identify asset $j$ in class $i$.
Denote the daily closing price for this asset as $S_{ij}(t)$ and it's corresponding log-return as $r_{ij}(t) = \ln \left(S_{ij}(t)/S_{ij}(t-1)\right)$.
For brevity, we will suppress the time variable for most of the discussion in this section.
Let $w_{ij}^{(p)}$ denote the weight of asset $ij$ in portfolio $p$, and $w_{ij}^{(b)}$ denote its weight in the benchmark.
We assume all weights are non-negative; that is, all portfolios considered take long-only positions.
Let $w_i^{(p)} = \sum_{j = 1}^{n_i}w_{ij}^{(p)}$ and $w_i^{(b)} = \sum_{j = 1}^{n_i}w_{ij}^{(b)}$ represent the total weights of the assets in class $i$
in the portfolio and benchmark respectively.
The quantities AA and SE for asset class $i$ are defined as follows:\footnote{See Chapter 5 in \cite{Bacon2008}.}
\begin{align}
\textrm{AA}_i &= \left(w_i^{(p)}-w_i^{(b)}\right)\left(R_i^{(b)}-R^{(b)}\right),\label{eq_AAi}
\\
\textrm{SE}_i &= w_i^{(b)}\left(R_i^{(p)}-R_i^{(b)}\right),\label{eq_SEi}
\end{align}
where
\begin{equation}
R_i^{(p)} = \sum_{j=1}^{n_i}\frac{w_{ij}^{(p)}}{w_i^{(p)}} \mathbb{E}\left[r_{ij}\right], \quad		
R_i^{(b)} = \sum_{j=1}^{n_i}\frac{w_{ij}^{(b)}}{w_i^{(b)}} \mathbb{E}\left[r_{ij}\right], \quad	
R^{(b)}   =  \sum_{i = 1}^{M}\sum_{j=1}^{n_i}w_{ij}^{(b)} \mathbb{E}\left[r_{ij} \right], \label{eq_R}
\end{equation}
and $\mathbb{E}\left[ \cdot \right]$ denotes expected value.
In (\ref{eq_R}), the ratio $w_{ij}^{(p)}/w_i^{(p)}$ represents the fractional weight held by asset $j$ in class $i$ in portfolio $p$.
(That is $\sum_{j=1}^{n_i} \left( w_{ij}^{(p)}/w_i^{(p)} \right)=1$.)
Thus $R_i^{(p)}$ (similarly $R_i^{(b)}$) represents an expected log-return for asset class $i$ consided as a fully-invested portfolio by itself.
In contrast, $R^{(b)}$ represents the usual expected log-return for the entire benchmark portfolio.\footnote{
 	If  $R_i^{(p)}, R_i^{(b)}, R^{(b)}$ and $r_{ij}$ were simple (i.e. discrete) returns, formulas of the form (\ref{eq_R}) are exact.
 	However, as they are log-returns, such formulas are approximate.
 	For example, the formula for $R^{(b)}$ incurrs an error term which, to leading order in a Taylor series  expansion, is
 	 $\frac{1}{2} \left[ \sum_{i=1}^M\sum_{j=1}^{n_i} w_{ij}^{(b)} \mathbb{E}\left[r_{ij} \right]^2
 	  -  \left( \sum_{i=1}^M\sum_{j=1}^{n_i} w_{ij}^{(b)} \mathbb{E}\left[r_{ij} \right] \right)^2 \right]$.
 	  \label{fn_logr}}
From (\ref{eq_R}) we have $R^{(b)} = \sum_{i = 1}^{M} w_i^{(b)} R_i^{(b)}$.
Similarly we have the usual expected log-return for portfolio $p$,
\begin{equation}
R^{(p)} = \sum_{i = 1}^{M}\sum_{j=1}^{n_i}w_{ij}^{(p)}\mathbb{E} \left[ r_{ij} \right] = \sum_{i = 1}^{M} w_i^{(p)} R_i^{(p)}. \label{eq_Rp}
\end{equation}

The excess return, $S = R^{(p)}-R^{(b)}$, can be viewed as the value added by portfolio management.
From (\ref{eq_AAi}) through (\ref{eq_Rp}),
\begin{equation}
S = \sum_{i = 1}^{M} \left( \textrm{AA}_i + \textrm{SE}_i + \textrm{I}_i \right) = \textrm{AA} +\textrm{SE} + \textrm{I}, \label{eq_S}
\end{equation}
where $\textrm{I}_i=\left(w_i^{(p)}-w_i^{(b)}\right)\left(R_i^{(p)}-R_i^{(b)}\right)$ is an ``interaction'' term.
AA, SE and I are, respectively, the total asset allocation, total selection effect, and total interaction terms for portfolio $p$.
The contribution to the total value added to the excess return, $S$, from asset class $i$ is $\textrm{AA}_i$,
while $\textrm{SE}_i$ represents the contribution to  $S$ determined by the choice of assets within class $i$.
To understand these interpretations, consider first the sign of the value of $\textrm{AA}_i$ in (\ref{eq_AAi}).
\begin{itemize}[leftmargin=10pt]
\item
If $R_i^{(b)}- R^{(b)} > 0$, the expected return from asset class $i$ in the benchmark is outperforming the total expected return for the benchmark.
Therefore if $w_i^{(p)}-w_i^{(b)} > 0$, the weight of asset class $i$ in portfolio $p$ is larger than in the benchmark, capitalizing further on the better return from class $i$.
Otherwise, if $w_i^{(p)}-w_i^{(b)} < 0$, the class $i$ weighting in portfolio $p$ is hurting the potential performance of that class (as determined by the benchmark).
\item
If $R_i^{(b)}- R^{(b)} < 0$, the expected return from asset class $i$ in the benchmark is under-performing the total expected return for the benchmark.
Therefore if $w_i^{(p)}-w_i^{(b)} < 0$, the weight of asset class $i$ in portfolio $p$ is smaller than in the benchmark, further suppressing the poorer return from that class.
Otherwise, if $w_i^{(p)}-w_i^{(b)} > 0$, the class $i$ weighting in portfolio $p$ is overweighting the poor performance of that class.
\end{itemize}
Thus a positive sign for the value of $\textrm{AA}_i$ indicates a ``correct'' decision in the management of portfolio $p$ relative to the benchmark
while a negative sign indicates a ``poor'' decision.
The magnitude of $\textrm{AA}_i$ quantifies how good or poor the decision is.

Similarly, as we assume\footnote{
	Also a requirement for class $i$ to be in the portfolio.}
$w_i^{(b)} > 0; \  i = 1, \dots, M$, a positive sign for the value of $\textrm{SE}_i$ in (\ref{eq_SEi}) indicates
that the expected return from the choice of assets in class $i$ in portfolio $p$ is outperforming that class in the benchmark,
while a negative sign indicates that the expected return from the choice of assets in class $i$ in portfolio $p$ is under-performing.

The interaction term, $\textrm{I}_i$, captures the part of the excess return unexplained by asset allocation and selection effect.
Written as
\begin{equation}
\textrm{I}_i=\frac{ \textrm{AA}_i \  \textrm{SE}_i }{ w_i^{(b)} \left(R_i^{(b)}-R^{(b)}\right)  }, \label{eq_Ii2}
\end{equation}
it can be viewed as  the product of the asset allocation and selection effect contributions of class $i$ to portfolio $p$
compared to the weighted excess return of class $i$ in the benchmark $b$.
Alternatively, written as
\begin{equation}
\textrm{I}_i = \left( \frac{w_i^{(p)}} {w_i^{(b)}} - 1 \right)\ \textrm{SE}_i, \label{eq_Ii}
\end{equation}
it can be interpreted as the product of the asset selection effect and the over- or under-weighted part of asset class $i$.
The  relationship (\ref{eq_Ii}) between $\textrm{I}_i$ and $\textrm{SE}_i$ reveals a simple form  for the sum of the selection effect and interaction terms,
\begin{equation}
\overline{\textrm{SE}}_i := \textrm{SE}_i + \textrm{I}_i = w_i^{(p)}\left(R_i^{(p)}-R^{(b)}\right). \label{eq_SEI}
\end{equation}
In the portfolio optimizations discussed next, (\ref{eq_SEI}) provides a way to incorporate a constraint on the sum of the selection and interaction effects
for class $i$.
We denote the total value of the combined selection effect plus interaction term by $\overline{\textrm{SE}} = \sum_{i=1}^M \overline{\textrm{SE}}_i$.

Expected tail loss (ETL), also known as conditional value-at-risk (CVaR), is defined in terms of value-at-risk (VaR).
Let $F(x) = Pr\{r\leq x\}$ denote the cumulative distribution function of a return $r$.
Then
\begin{align}
\textrm{VaR}_{\alpha}(r) &= -\inf\left\{ x \in \mathbb{R} \mid F(x) \ge 1-\alpha ,\right\}, \label{eq_VaR}	
\\
\textrm{ETL}_{\alpha}(r) &= -\mathbb{E}\left[ r | r \leq -\textrm{VaR}_{\alpha}(r) \right], \label{eq_ETL}
\end{align}
where ${\alpha} \in (0,1)$ is a prescribed quantile level, typically having value of 0.95, 0.99, or 0.995.
We consider four portfolio optimization problems, P$^k_{\alpha}, k = 0, \dots, 3$, based upon constrained minimization of ETL$_{\alpha}$ \citep{Krokhmal2002}.
These optimization problems successively add further performance attribute constraints to a long-only, fully invested, ETL$_{\alpha}$-minimized portfolio.
For the total portfolio return $R^{(p)}$ (\ref{eq_Rp}), perform the minimization
\begin{equation}
\min_{w_{ij}^{(p)}} \textrm{ETL}_{\alpha} \left( R^{(p)} \right), \label{eq_opt}
\end{equation}
subject to the constraints:
\begin{itemize}[leftmargin=20pt]
\item[P$^0_{\alpha}$:]
(a) $w_{ij}^{(p)}\geq 0,\ \sum_{i,j}w_{ij}^{(p)}= 1$.
\item[P$^1_{\alpha}$:]
(a) $w_{ij}^{(p)}\geq 0,\ \sum_{i,j}w_{ij}^{(p)}= 1$; and (b) $a_1\leq \textrm{AA} \leq b_1$.
\item[P$^2_{\alpha}$:]
(a) $w_{ij}^{(p)}\geq 0,\ \sum_{i,j}w_{ij}^{(p)}= 1$; (b) $a_1\leq \textrm{AA} \leq b_1$; and (c) $a_2\leq \textrm{SE}\leq b_2$.
\item[P$^3_{\alpha}$:]
(a) $w_{ij}^{(p)}\geq 0,\ \sum_{i,j}w_{ij}^{(p)}= 1$; (b) $a_1\leq \textrm{AA} \leq b_1$; and (c) $a_3\leq \overline{\textrm{SE}}\leq b_3$.
\end{itemize}
The constants $a_i ,b_i$ can be user-specified to meet particular goals.
For example, the constraint $\textrm{AA} \ge 0$ requires that, on average, the asset classes in the optimized portfolio $p$ equal-or-outperform
those in the benchmark.
A constraint $\textrm{SE}_i \ge 0$ requires that the weights of the assets in class $i$ be adjusted to perform as well, or better than, class $i$ in the benchmark.
Since individual asset weights can be zero, this is equivalent to choice of assets in the class.
The constraint $\textrm{SE} \ge 0$ requires that this be true averaged over classes.
We note that the minimization (\ref{eq_opt}) can be solved in terms of a linear optimization function \citep{Tutuncu2003}.
As $R_i^{(p)}(t)$ involves the ratio $w_{ij}^{(p)}(t)/w_i^{(p)}(t)$, constraints involving $\textrm{SE}_i$ terms are non-linear.
In contrast, constraints involving terms $\textrm{AA}_i$ and $\overline{\textrm{SE}}_i$ are linear.\footnote{
	This assumes that benchmark weight values can be obtained in a timely time and are not part of the optimization.}

Performance of these four optimized portfolios, relative to each other and to the benchmark, will be judged based upon cumulative price,
as well as performance relative to three common risk measures.
Let $w_{ij}^{(p)}(t)$, $i = 1,\ldots,M$, $j = 1,\ldots,n_i$, $t = 1,\ldots,T$ denote daily weights obtained from one of these optimizations.\footnote{
	Specifially $w_{ij}^{(p)}(t)$ is the optimized weight to be applied to the portfolio at the beginning of (and throughout the entire)
	 day $t$.}
Recalling that $r_{ij}(t)$ is the log-return based upon the closing price of asset $ij$ on day $t$,
the portfolio log-return$^{\ref{fn_logr}}$ and cumulative price are
 $R^{(p)}(t) = \sum_{i=1}^M\sum_{j=1}^{n_i} w_{ij}^{(p)}(t) r_{ij}(t)$
 and
 $S^{(p)}(t) = S_0\exp\left (\sum_{s=1}^{t} R^{(p)}(s) \right)$.
The three measures used are:
\begin{itemize}[leftmargin=15pt]
	\item [1.]
	maximum drawdown (MDD),
	\begin{equation*}
		\textrm{MDD}(T) = \sup_{t\in [0,T]}\left[ \sup_{s \in [0,t]} (S^{(p)}(s)-S^{(p)}(t) ) \right],
	\end{equation*}
	 which characterizes the maximum loss incurred from peak to trough during the time period $[0,T]$;
	\item [2.]
	Sharpe ratio \citep{Sharpe1994},
	\begin{equation*}
		\textrm{Sharpe}(T) = \frac{ \mathbb{E}[R^{(p)}(t) - r_f(t)]_{[0,T]} } {\sqrt { \text{var}[ R^{(p)}(t) - r_f (t)]_{[0,T]} } }
			 = \frac{ \mu_p |_{[0,T]} } { \sigma_p |_{[0,T]} },
	\end{equation*}
	where $r_f(t)$ is a risk-free rate,
	and $\mu_p$ and $\sigma_p$ are the expected mean and standard deviation of the portfolio's excess return, $R^{(p)}(t) - r_f(t)$;
	and
	\item [3.]
	Rachev ratio \citep{Rachev2008},
	\begin{equation*}
		\textrm{Rachev}_{\alpha,\beta}(T) = \frac{ \textrm{ETL}_{\alpha} (r_f(t) - R^{(p)}(t)) } { \textrm{ETL}_{\beta} (R^{(p)}(t) - r_f(t)) },
	\end{equation*}
	which represents the reward potential for positive returns compared to the risk potential for negative returns at quantile levels defined by the user.
	In our analysis, we set $\alpha=\beta=0.95$.
\end{itemize}

%***************************************************************************

\section{Application to a Test Portfolio}
\label{sec3}
\noindent
To illustrate portfolio optimization under performance attribution constraints,
we consider a specific portfolio comprised of stocks from the Dow Jones Industrial Average (DJIA).
As a limited-information index of the performance of the U.S. stock market,
the DJIA consists of the weighted stock price of 30 large, publicly-traded companies.
The stock composition of the DJIA and their weights in the index, as of 02/01/2021, are presented in Table \ref{tab1_DJIA} of the Appendix.
To preserve a sufficiently long trading history, our test portfolio comprises 29 of the 30 stocks from the DJIA.\footnote{
	We exclude Dow Inc. which was spun off of DowDuPont on April 1, 2019.
	Its stock, under the ticker symbol ``DOW'', began trading on March 20, 2019. It was added to the DJIA on April 2, 2019.}
Daily closing price data for all 29 stocks were available\footnote{
	Bloomberg Professional Services.}
covering the period 03/19/2008 through 02/01/2021.
We grouped the stocks in our test portfolio into six classes based upon their weighted value in the DJIA.
Class composition and their total weight in the DJIA are presented in Table \ref{tab_class}.
Our benchmark is the equi-weighted portfolio of the same 29 stocks, grouped in the same classes.\footnote{
	As is apparent from equations (\ref{eq_AAi}) through (\ref{eq_Ii}), results from an attribution analysis depend critically on choice of benchmark.}
The 10-year U.S. Treasury yield curve rate was used as the risk-free rate.

\begin{table}[ht]
	\begin{center}
		\caption{Breakdown of the asset classes in the test portfolio}
		\label{tab_class}
		\begin{tabular}{c l c}
			\textrm{Class}&{Stock Ticker}&{Class Weight in DJIA(\%)}\\
			\midrule		
			\textrm{1}&\textrm{UNH, GS, HD, AMGN, MSFT}&$29.59$\\
			\textrm{2}&\textrm{CRM, MCD, V, BA, HON}&$22.35$\\
			\textrm{3}&\textrm{CAT, MMM, DIS, JNJ, WMT}&$18.13$\\
			\textrm{4}&\textrm{TRV, NKE, AAPL, JPM, PG}&$14.52$\\
			\textrm{5}&\textrm{IBM , AXP, CVX, MRK, INTC}&$9.97$\\
			\textrm{6}&\textrm{VZ, WBA, KO, CSCO}&$4.29$\\
		\end{tabular}
	\end{center}
\end{table}

Daily return data for the stocks covered $3,240$ trading days.
Using a standard rolling-window strategy for optimization with a window size of 1,008 days (four years),
optimized portfolio return data were computed for an in-sample period of $T = 2,232$ days.
In the portfolio optimization, we assume no transaction costs and impose no turnover constraints.
We optimized at two separate quantile levels, $\alpha \in \{0.95,0.99\}$.
For the attribution constraints in optimizations P$^1_{\alpha}$, P$^2_{\alpha}$ , and P$^3_{\alpha}$,
we set the lower bounds  $a_i = 0, i = 1, 2, 3$, and set no upper bounds ($b_i = \infty, i = 1, 2, 3$).
Thus, for example, optimization P$^1_{\alpha}$ minimizes the ETL for the long-only portfolio while requiring that, on average, its asset classes
outperform the benchmark.

To circumvent the non-linear SE constraints in portfolio optimization P${}_\alpha^2$,
we performed a two-step optimization procedure to linearize the SE constraints.\footnote{
	The two-step optimization was employed only for optimizations P${}_\alpha^2$, P${}_\alpha^5$, P${}_\alpha^7$ and P${}_\alpha^8$.}
The first step satisfies condition (b) of optimization P$^2_{\alpha}$ by performing optimization P$^1_{\alpha}$ to provide the optimal class weights $w_{ij}^{(p)}$.
The second step satisfies condition (c) using the fixed class weights from the first step.

\begin{figure}[ht]
    \begin{center}
        \begin{subfigure}[b]{0.6\textwidth} 
		\includegraphics[width=\textwidth]{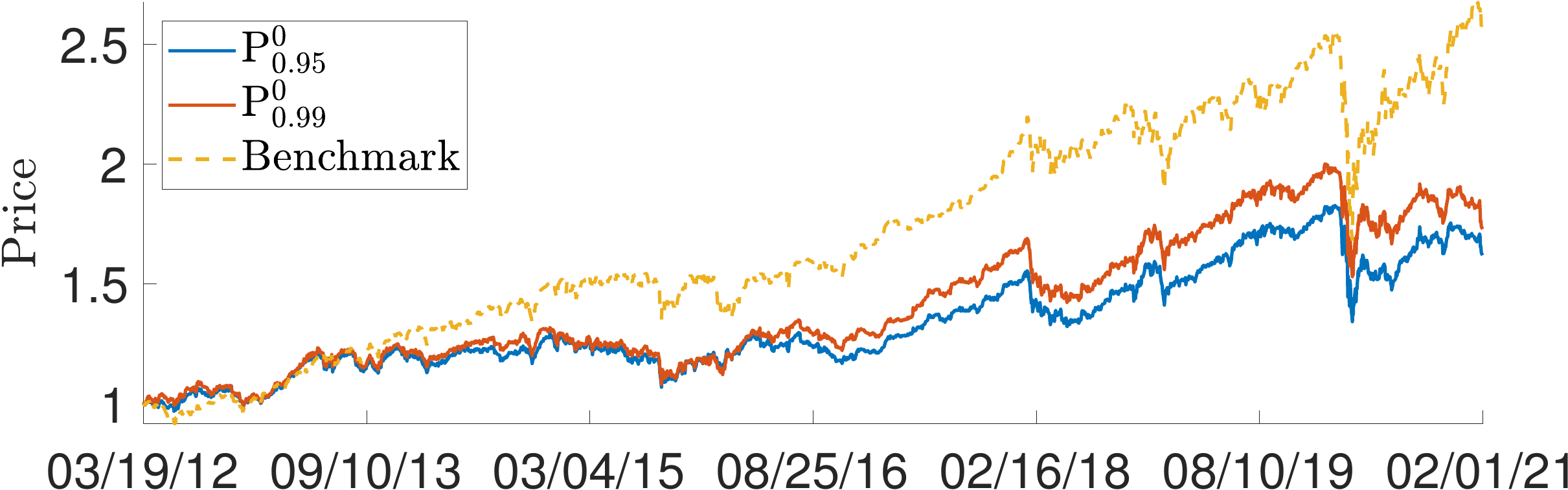}
	\end{subfigure}
	\hfill%
	\begin{subfigure}[b]{0.35\textwidth}
		\includegraphics[width=\textwidth]{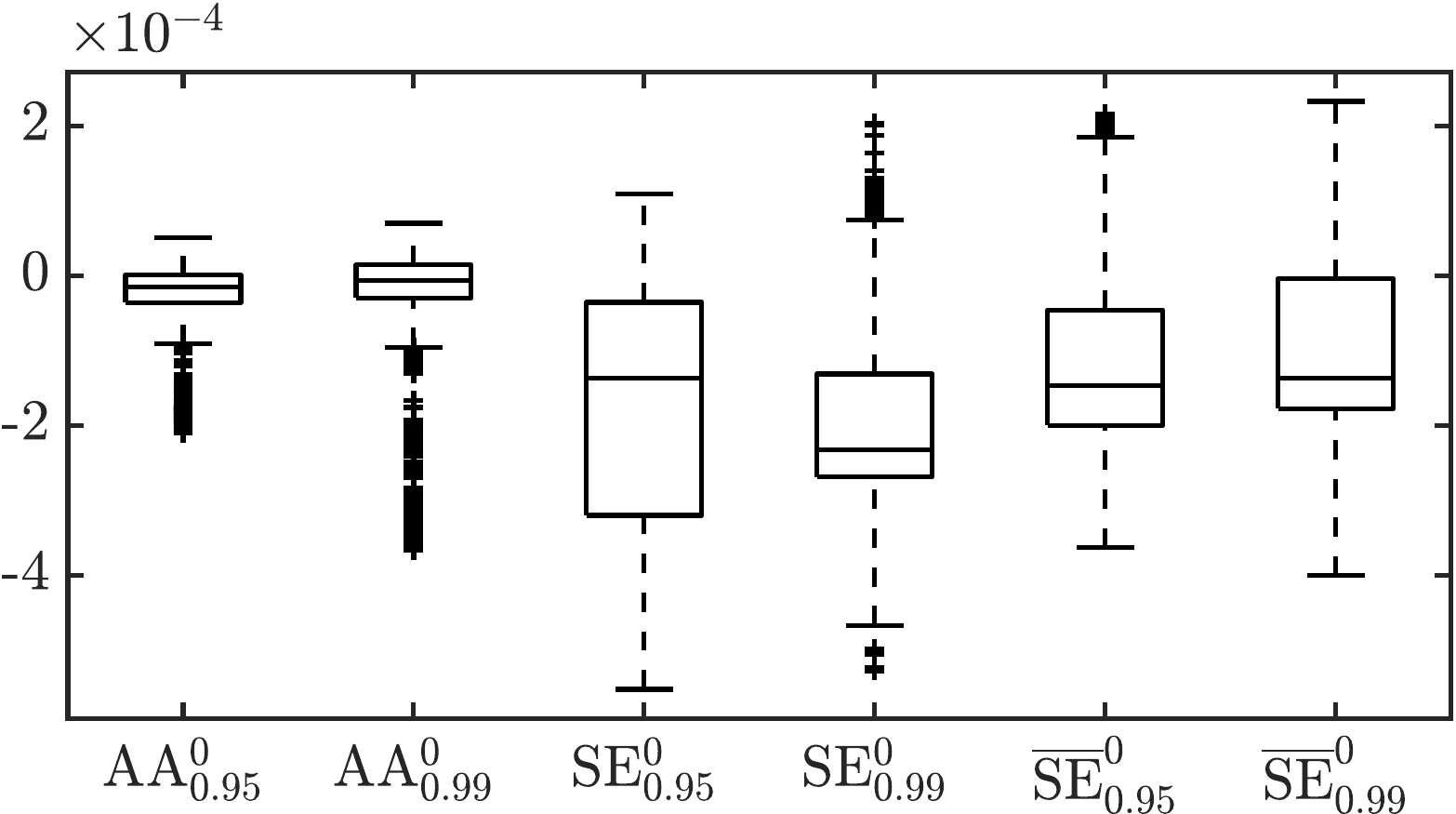} 
	\end{subfigure}
	\begin{subfigure}[b]{0.6\textwidth} 
		\includegraphics[width=\textwidth]{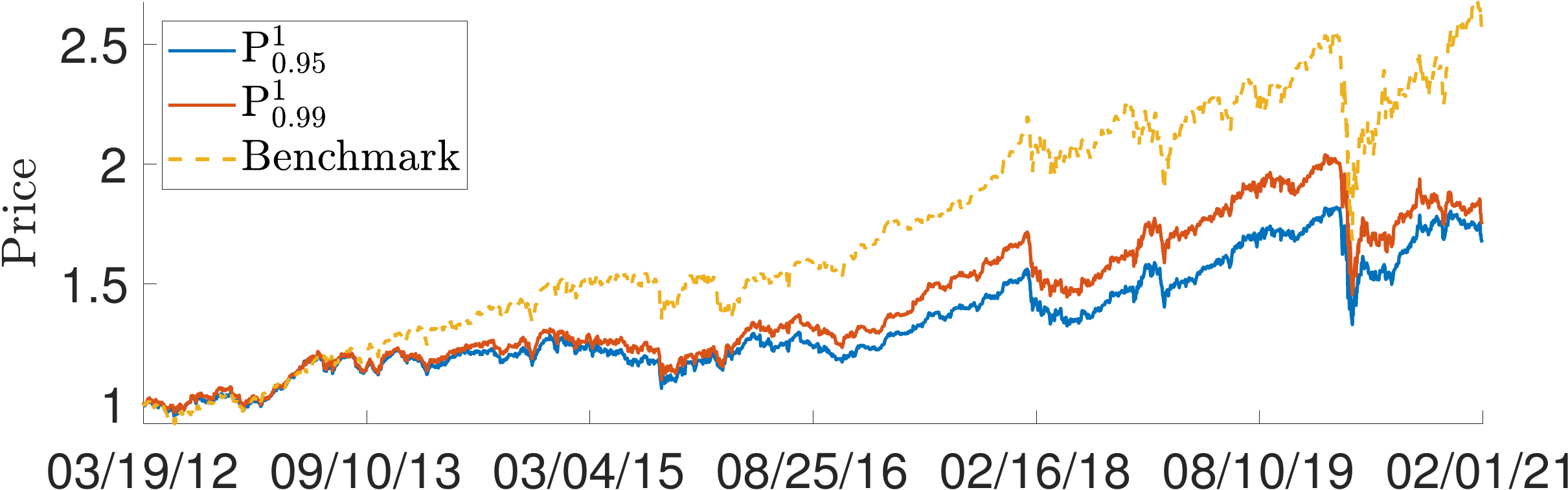}
	\end{subfigure}
	\hfill%
	 \begin{subfigure}[b]{0.35\textwidth} 
		\includegraphics[width=\textwidth]{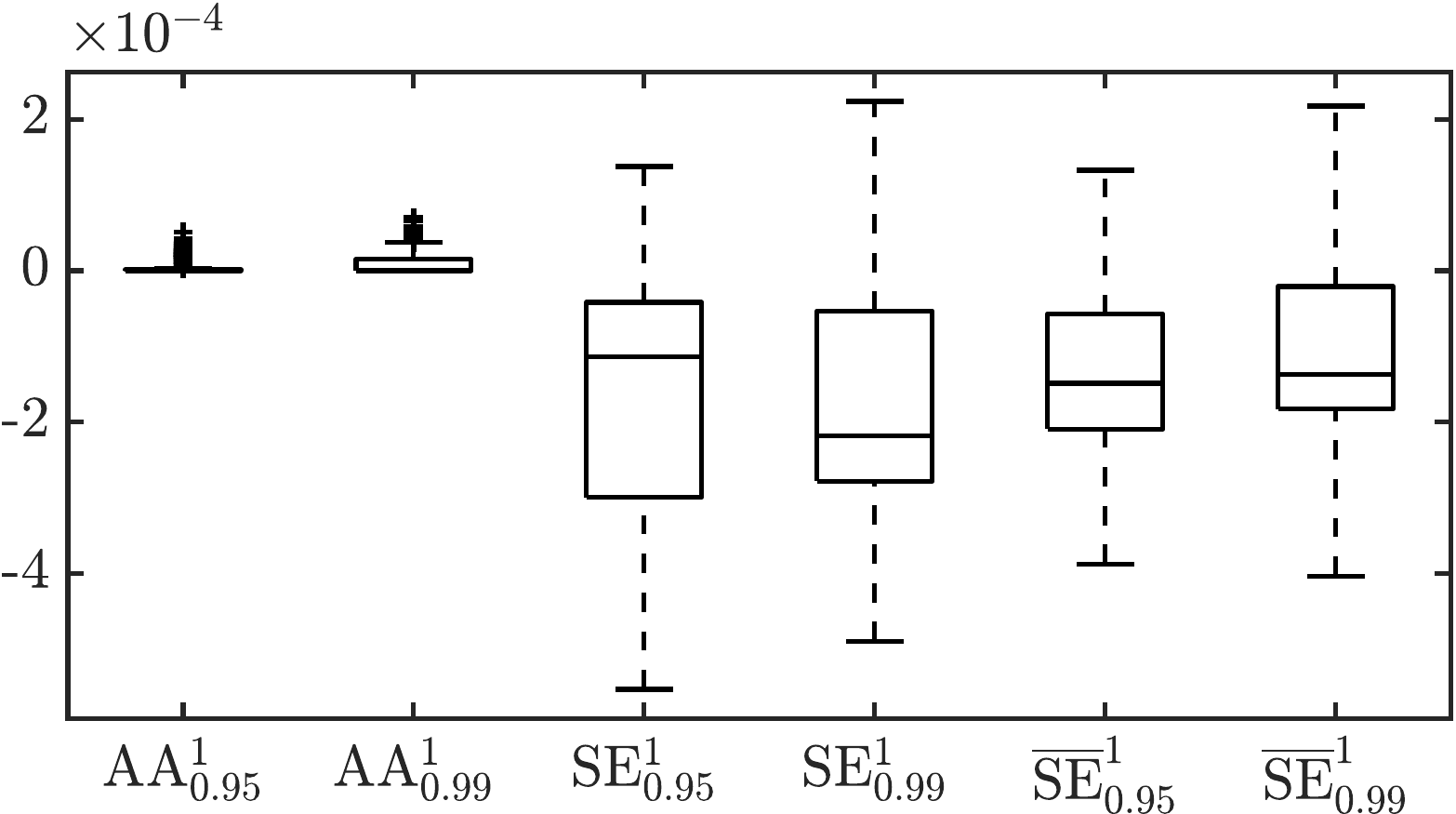}
	\end{subfigure}
	\begin{subfigure}[b]{0.6\textwidth} 
		\includegraphics[width=\textwidth]{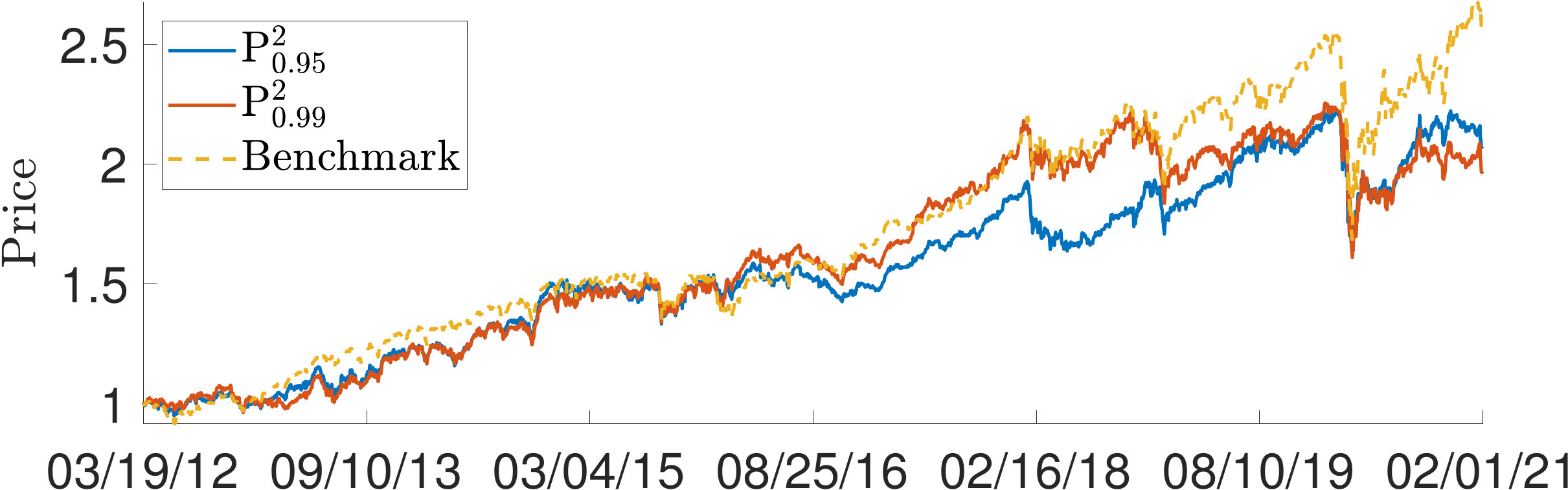}
	\end{subfigure}
	\hfill%
	\begin{subfigure}[b]{0.38\textwidth} 
		\includegraphics[width=\textwidth]{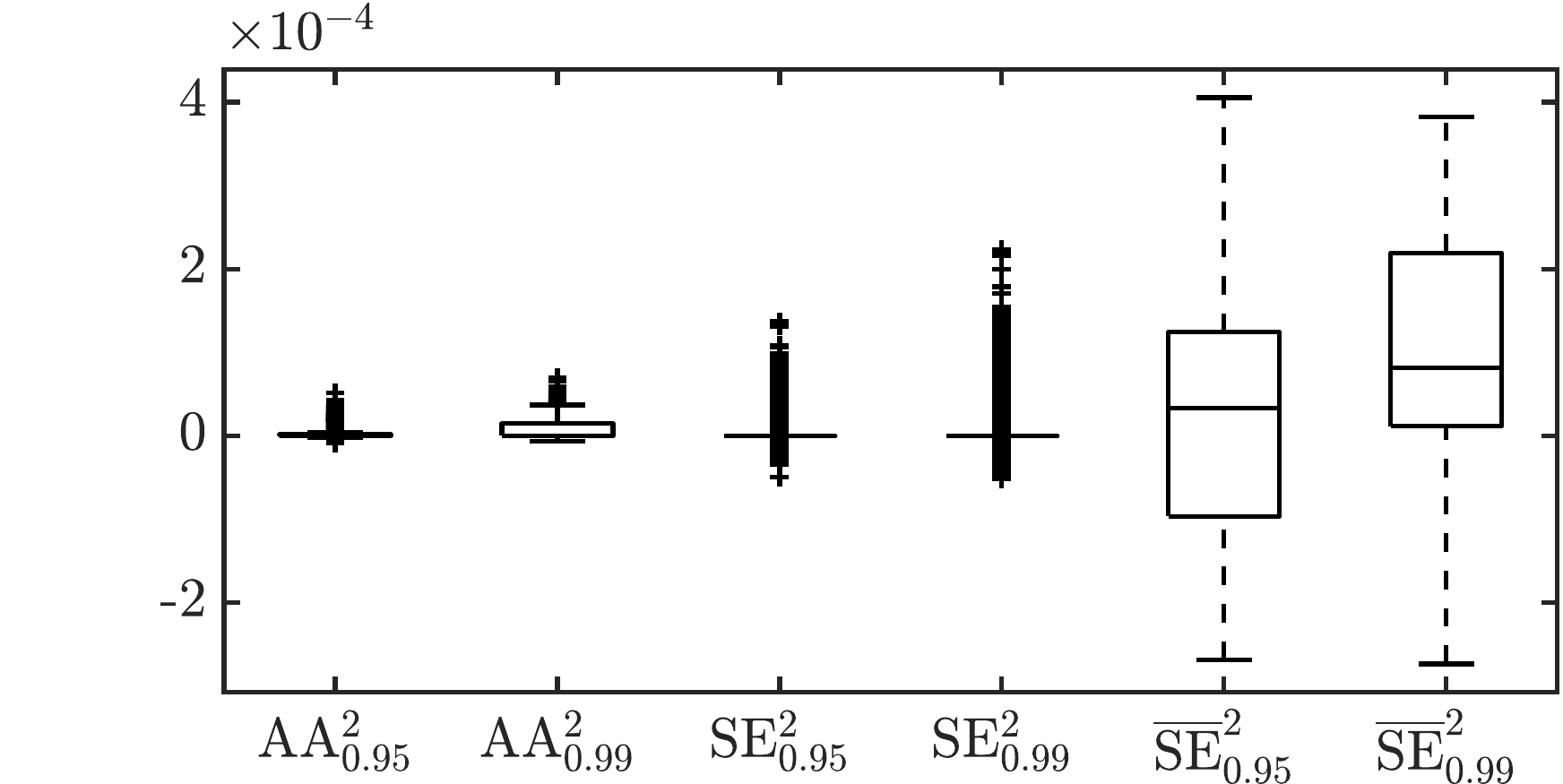}
	\end{subfigure}
	\begin{subfigure}[b]{0.6\textwidth} 
		\includegraphics[width=\textwidth]{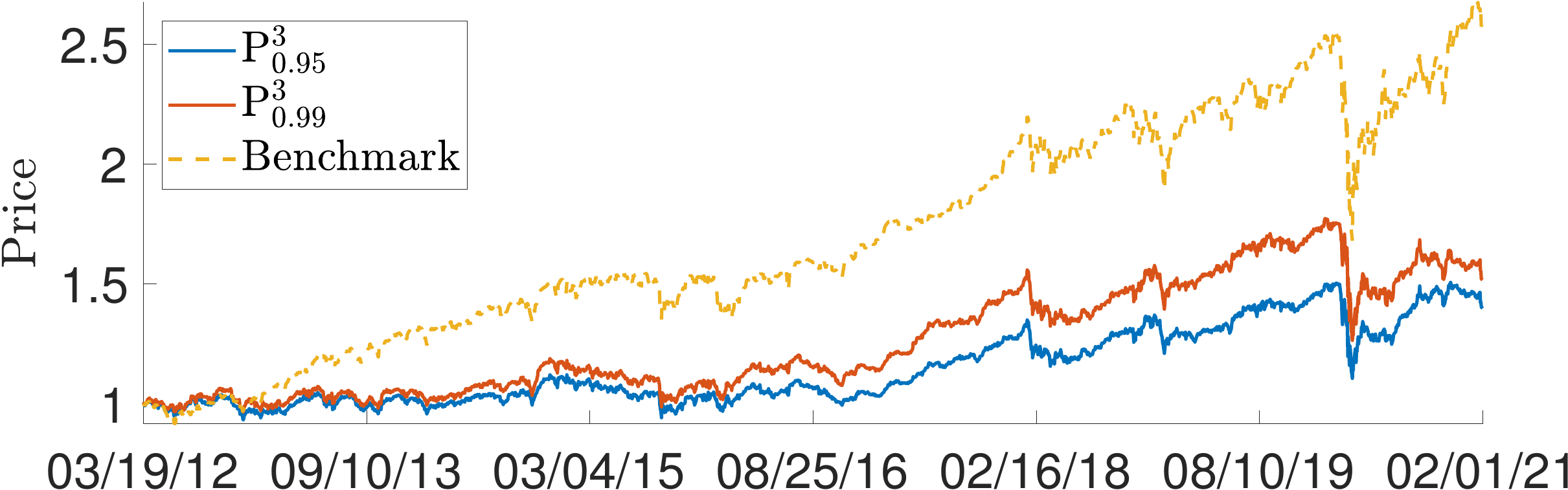}
		\caption{}
		\label{fig_P0123_price}
	\end{subfigure}
	\hfill%
	\begin{subfigure}[b]{0.38\textwidth} 
		\includegraphics[width=\textwidth]{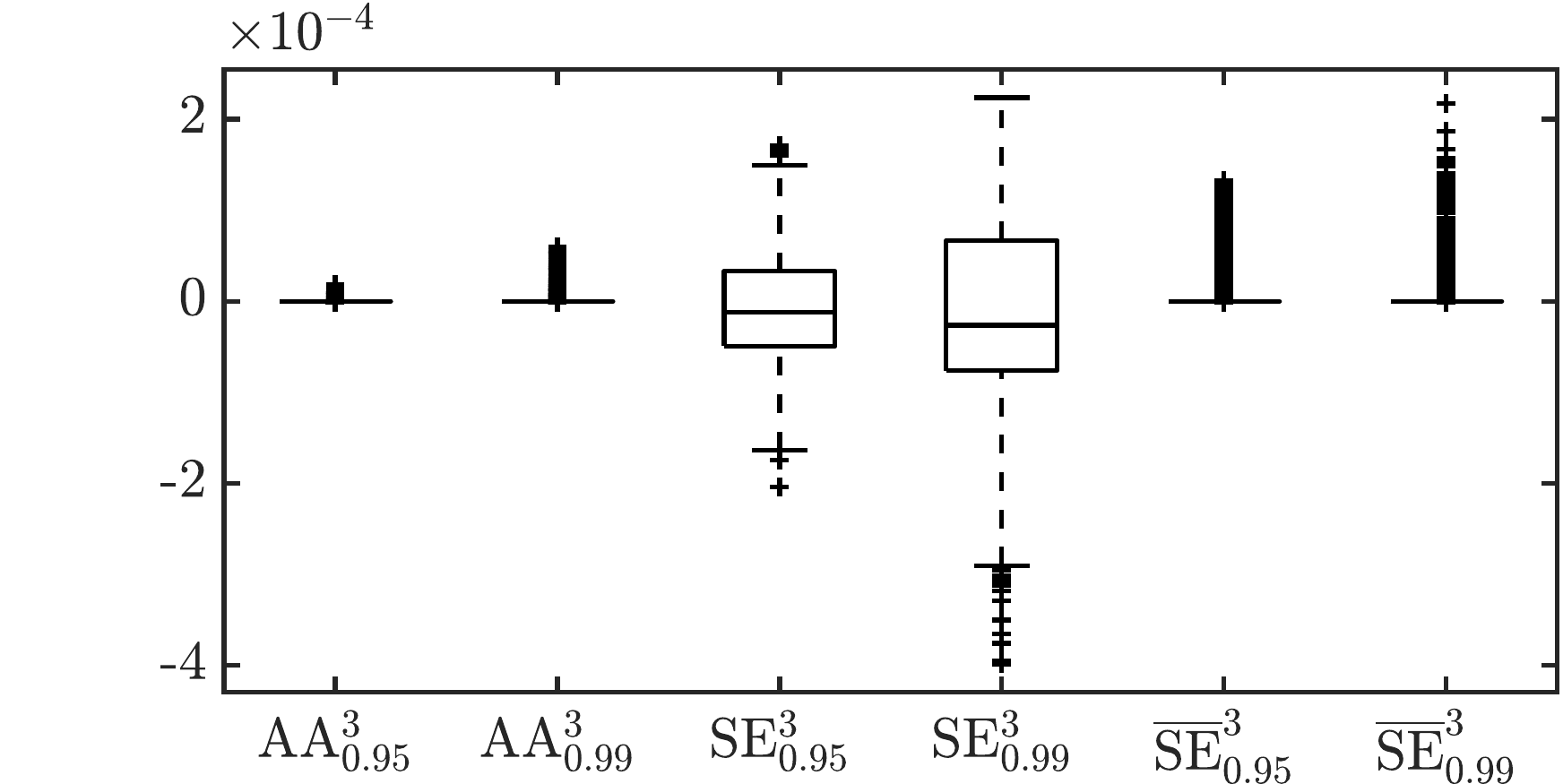}
		\caption{} 
		\label{fig_P0123_box}
	\end{subfigure}
	\caption{(a) Cumulative price resulting from a \$1 investment and
		     (b) box-whisker plots of the observed distributions of AA, SE and $\overline{\textrm{SE}}$ values
 			for the optimized portfolios P${}_\alpha^0$ through P${}_\alpha^3$.}
    \label{fig_P0123}
    \end{center}
\end{figure}
Fig. \ref{fig_P0123_price} presents comparisons of the cumulative price performance for the benchmark and
the optimized portfolios P${}_\alpha^0$ through P${}_\alpha^3$.
Appendix Fig.~\ref{fig_Sij} provides comparative cumulative price plots for the individual assets in the portfolio.
Fig. \ref{fig_P0123_box} summarizes the statistics on the distributions of the daily values of AA, SE and $\overline{\textrm{SE}}$
for the optimized portfolios.
In this figure, in the text and in similar figures below,
we employ the notation AA$^0_{0.95}$ to refer to AA computed for P$^0_{0.05}$, etc.

\begin{figure}[h]
	\begin{center}
		\includegraphics[width=0.65\textwidth]{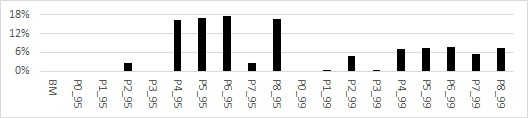}
		\caption{Percent of optimizations resulting in no weight solution in the feasibility region.
			(BM = benchmark; P0\_95 = P$^0_{0.95}$, etc.)}
		\label{fig_FR}
	\end{center}
\end{figure}

\begin{itemize}[leftmargin=10pt]
\item
As expected, compared to the benchmark the optimized portfolios P${}_\alpha^0$ develop a less aggressive overall price growth by reducing ETL,
with P$_{0.99}^0$ providing better cumulative return than P$_{0.95}^0$.
Reflecting the decreased return performance of P$_\alpha^0$ relative to the benchmark,
values of AA, SE and $\overline{\textrm{SE}}$ are overwhelming negative,
from 58\% of the daily values being negative for AA$^0_{0.99}$ to 93\% for SE$^0_{0.99}$.
The spread of values is smallest for AA and much wider for SE and $\overline{\textrm{SE}}$.
\item
For the same value of $\alpha$, differences between the price processes for P${}_\alpha^1$ and P${}_\alpha^0$ are
relatively minor.
Most noticeable is a narrowing of the difference between the performance of
P${}_{0.95}^1$ and P${}_{0.99}^1$ in the pandemic period.
The narrow distribution of AA${}_\alpha^1$ values reflect the non-negative constraint;
For the daily values of AA${}_{0.95}^1$, 68\% of  are zero, while 54\% are zero for AA${}_{0.99}^1$.
Compared to P$^0_\alpha$, there are minor changes in the distributions of SE$^1_\alpha$ and $\overline{\textrm{SE}}^1_\alpha$;
the most notable difference being a widening (toward positive values) of the SE$^1_{0.99}$ distribution.
\item
Noticeable improvement occurred in the cumulative price performance of P$_\alpha^2$;
in particular portfolio P$_{0.99}^2$ tracks the benchmark until 12/06/2018.
The non-negative constraints placed on AA and SE are reflected in the distribution changes;
82\% of the daily values for SE${}_{0.95}^1$ are now zero, while 80\% are zero for SE${}_{0.99}^1$.
The negative outlier values seen in the SE${}_\alpha^1$ distributions occur as the result of not being able to
locate optimized solutions in the feasibility region every time-step.
(See Fig.~\ref{fig_FR} and related discussion.)
As values of $\overline{\textrm{SE}}$ are closely related to values of SE (see (\ref{eq_Ii})),
a strong shift towards positive values in the distributions of $\overline{\textrm{SE}}^2_\alpha$ occurred.
\item
The attempt to constrain AA and $\overline{\textrm{SE}}$ in optimizations P${}_\alpha^3$ resulted in decreased price performance, with essentially no cumulative price increase from 2012 through 2016.
The non-negative constraints on $\overline{\textrm{SE}}$ resulted in
85\% of the daily values for $\overline{\textrm{SE}}{}_{0.95}^3$ being zero;
with 79\% zero for $\overline{\textrm{SE}}{}_{0.99}^1$.
\end{itemize}

\begin{figure}[t]
	\begin{center}
		\includegraphics[width=0.65\textwidth]{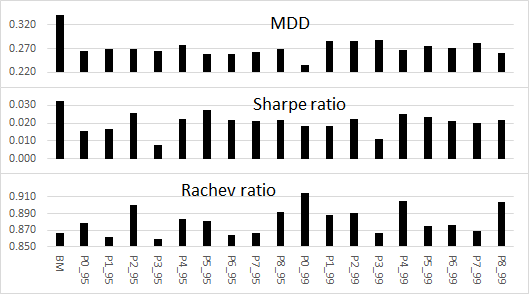}
		\caption{Total-time-period risk measures for the portfolios in the study. (BM = benchmark; P0\_95 = P$^0_{0.95}$, etc.)}
		\label{fig_RM}
	\end{center}
\end{figure}
Fig. \ref{fig_RM} provides the risk-measure values for the benchmark and these four optimized portfolios.
Each risk measure value represents the entire time period of study.
\begin{itemize}[leftmargin=10pt]
\item
Compared with a 34\% MDD for the benchmark,
these optimized portfolios reduce MDD by a further: $\sim$5\% for P$^k_{0.99}, k = 1,2,3$;
7.5\%  for P$^k_{0.95}, k = 0,1,2,3$; and 11\% for P$_{0.99}^0$.
\item
Compared with a value of 0.032 for the benchmark, Sharpe ratios for these optimized portfolios are reduced:
by a factor of $\sim$2 for P$^0_\alpha$ and P$^1_\alpha$; by a factor of 1.2 to 1.5 for  P$^2_\alpha$;
and by a factor of 3 to 4 for P$^3_\alpha$.
The large decrease in Sharpe ratio for optimizations P$^3_\alpha$ reflect their relatively flat price performance.
\item
Compared with the benchmark value of 0.867, Rachev ratios for these optimized portfolios generally increase (by 2\% to 5\%),
with the exceptions of P$^3_\alpha$ and P$^1_{0.95}$.
\item
The improved price performance of P$^2_\alpha$ (compared to the other optimized portfolios) is also accompanied by improved
values in MDD and Rachev ratio (compared to the benchmark).
\end{itemize}

\begin{figure}[h]
\begin{center}
        \begin{subfigure}[b]{0.49\textwidth} 
    	\includegraphics[width=\textwidth]{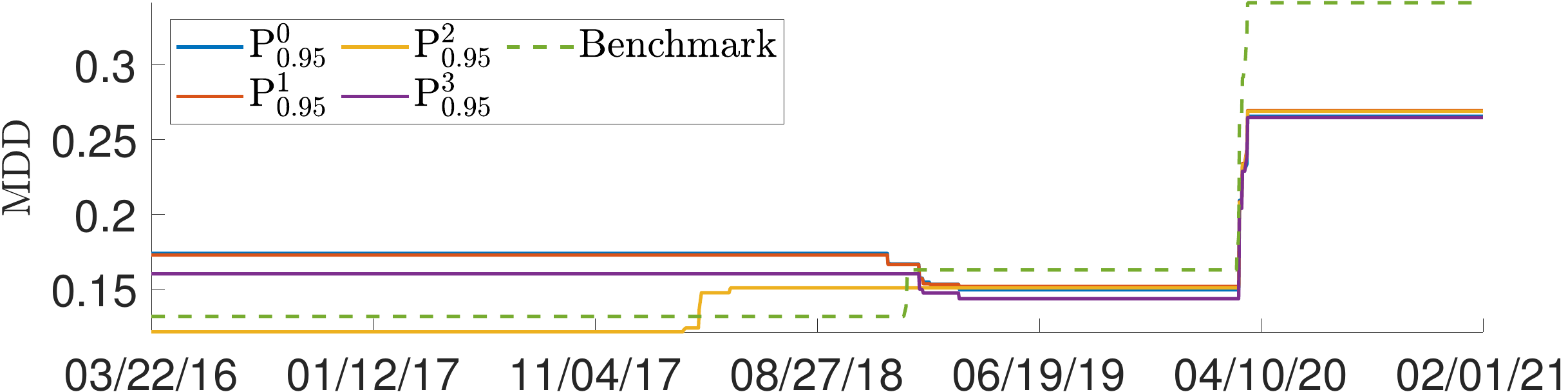}
    	\vspace{-.7cm}
    \end{subfigure}
            \begin{subfigure}[b]{0.49\textwidth} 
    	\includegraphics[width=\textwidth]{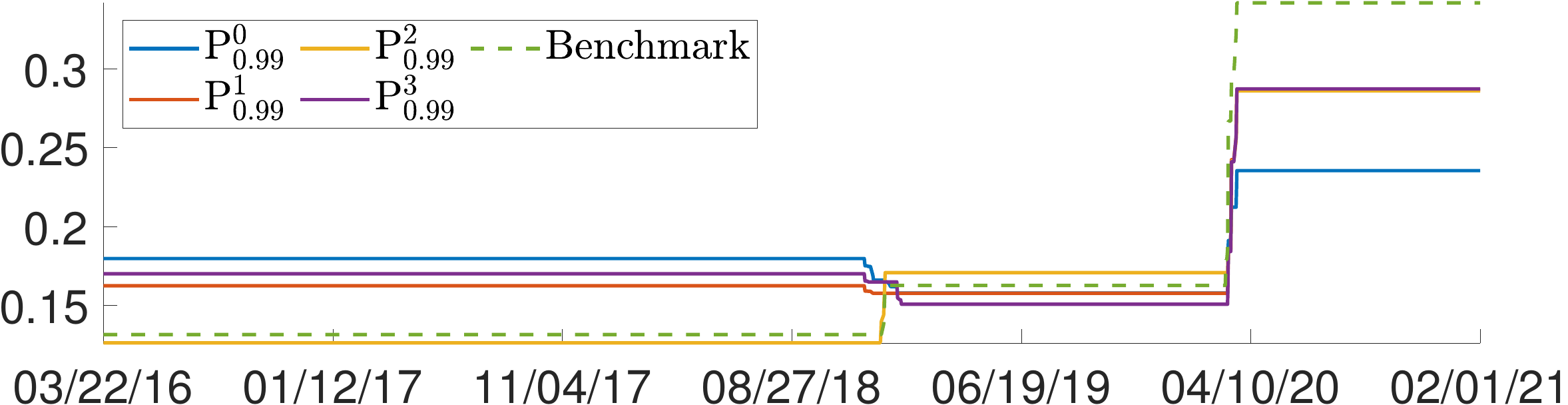}
    	\vspace{-.7cm}
    \end{subfigure}
    \begin{subfigure}[b]{0.49\textwidth} 
    	\includegraphics[width=\textwidth]{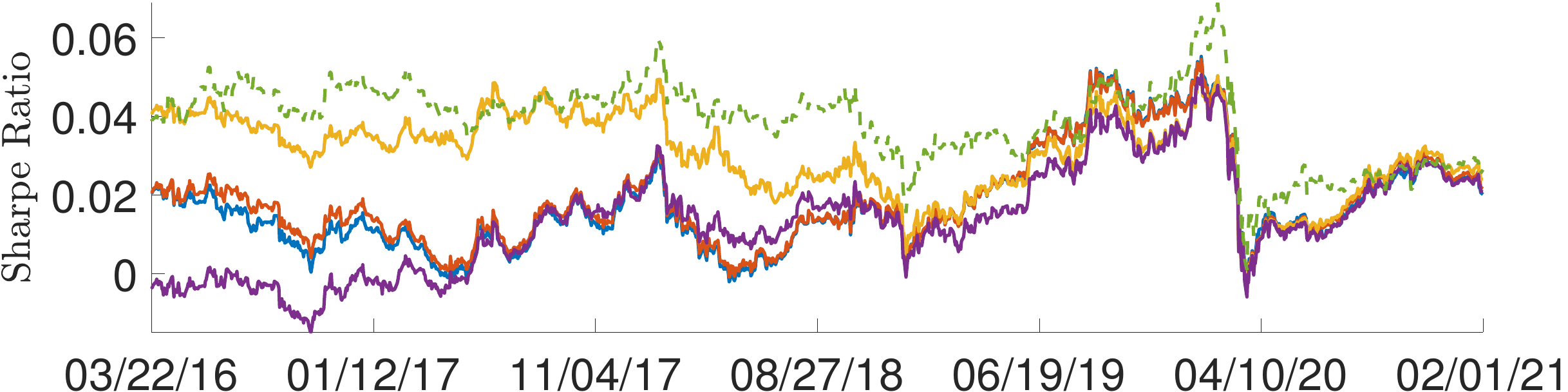}
    	\vspace{-.7cm}
    \end{subfigure}
        \begin{subfigure}[b]{0.49\textwidth} 
    	\includegraphics[width=\textwidth]{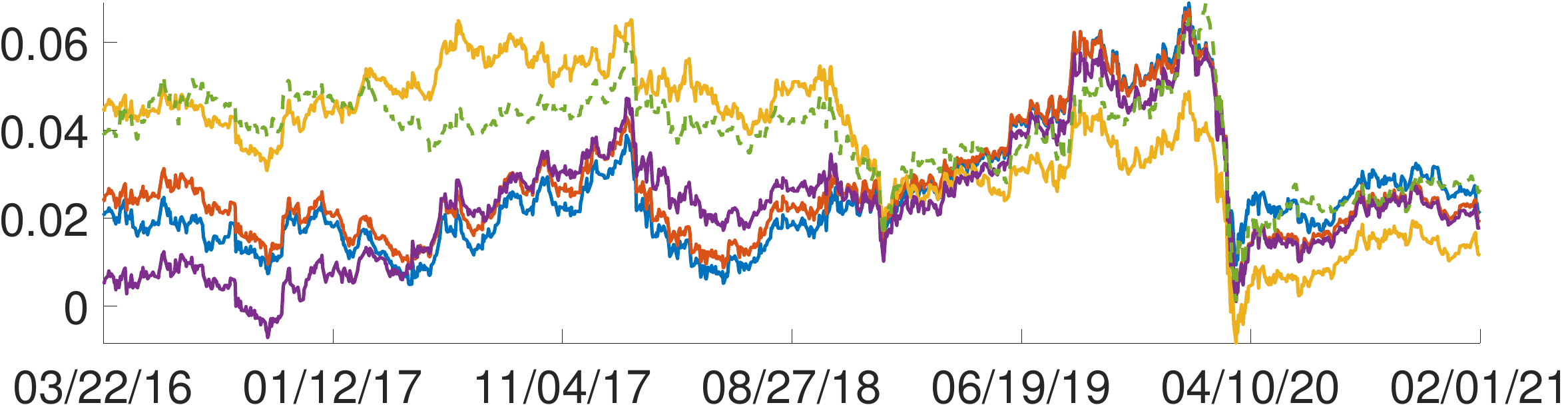}
    	\vspace{-.7cm}
    \end{subfigure}
    \begin{subfigure}[b]{0.49\textwidth} 
    	\includegraphics[width=\textwidth]{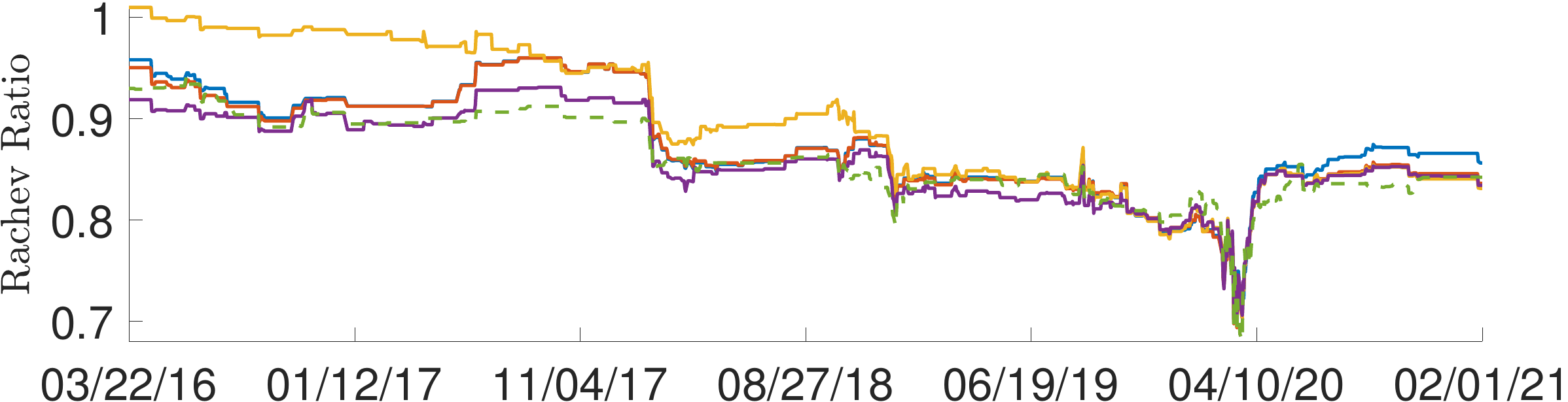}
    \end{subfigure}
        \begin{subfigure}[b]{0.49\textwidth} 
    	\includegraphics[width=\textwidth]{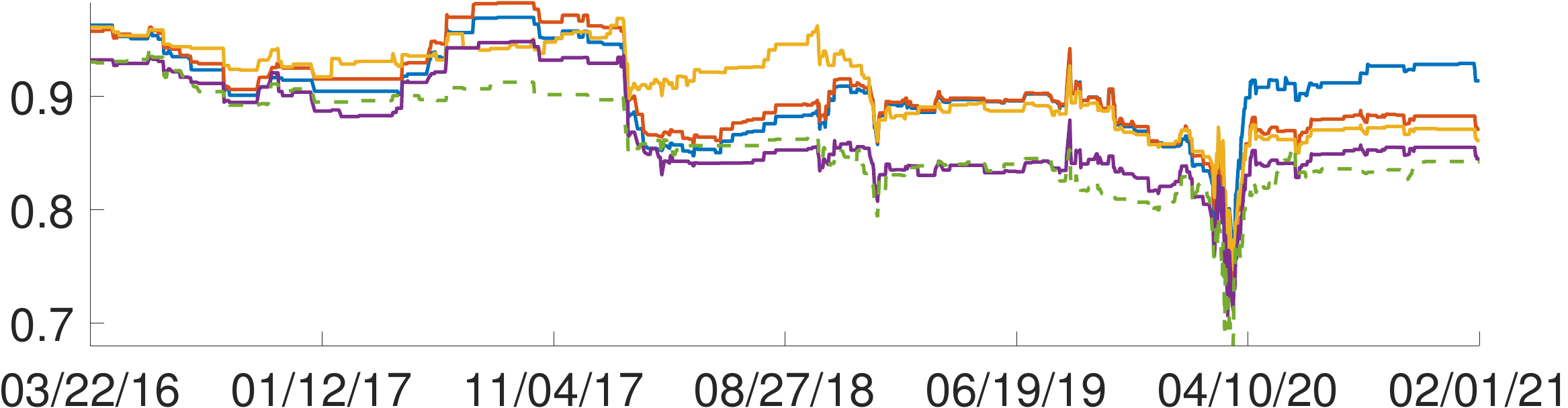}
    \end{subfigure}
    \caption{Risk measures computed using a 1,008-day moving window for the benchmark and the portfolios
		(left) P$^k_{0.95}$ and (right) P$^k_{0.99}, k = 0, \dots 3$.}
    \label{fig_RMmw}
\end{center}
\end{figure}
The risk-measure values presented in Fig. \ref{fig_RM} provide a single value for the entire in-sample data period.
One consequence is that the MDD values in Fig.~\ref{fig_RM} reflect only the onset of the Covid-19 pandemic.
Fig.~\ref{fig_RMmw} therefore plots daily risk measure values, computed using a 1,008-day moving window,
for the benchmark and P$^0_\alpha$ through P$^3_\alpha$ over the period 3/22/2016 through 02/01/2021.
The MDD data suggest three separate time periods, 2016:Q2 through 2018:Q4, 2019:Q1 though 2020:Q1, and 2020:Q2 into 2021:Q1.
In the first time period, the benchmark and P${}_{\alpha}^2$ display the lowest MDD while the benchmark has the highest MDD in period three.
The benchmark has the highest Sharpe ratio in period one, but the optimized portfolios, particularly those for $\alpha = 0.99$,
become competitive with the benchmark in periods two and three.
Generally the benchmark and P$^3_\alpha$ have the worst Rachev ratio.
Of these optimized portfolios, P${}_\alpha^2$ generally displays the best risk measures.
During the pandemic (period three), all P${}_{0.95}^k, k = 0,1,2,3$ portfolios display comparable risk measures.
There is greater variation in the performance of the P${}_{0.99}^k, k = 0,1,2,3$ portfolios during the pandemic period,
with P${}_{0.99}^0$ being a consistent best performer.

The addition of the performance attribution total measures AA, SE and $\overline{\textrm{SE}}$
as constraints do not result in uniformly improved performance relative to the ETL-optimized portfolio P${}_\alpha^0$.
To test whether this is the result of using total measures that only constrain averages over all classes,
we attempted a more aggressive set of optimizations which apply asset allocation and selection effect constraints to each individual class:
\begin{itemize}[leftmargin=20pt]
\item[P$^4_{\alpha}$:]
(a) $w_{ij}^{(p)}\geq 0,\ \sum_{i,j}w_{ij}^{(p)}= 1$; and (b) $0\leq \textrm{AA}_i,\ i = 1, \dots, M$.
\item[P$^5_{\alpha}$:]
(a) $w_{ij}^{(p)}\geq 0,\ \sum_{i,j}w_{ij}^{(p)}= 1$; (b) $0\leq \textrm{AA}_i$;
	and (c) $0\leq \textrm{SE}_i \textrm{ when } w_i^{(p)} > 0,\ i = 1, \dots, M $.
\item[P$^6_{\alpha}$:]
(a) $w_{ij}^{(p)}\geq 0,\ \sum_{i,j}w_{ij}^{(p)}= 1$; (b) $0\leq \textrm{AA}_i$; and (c) $0\leq \overline{\textrm{SE}}_i,\ i = 1, \dots, M$.
\end{itemize}

Constraint (c) for P$^5_\alpha$ requires explanation.
As with P$^2_\alpha$, to avoid the non-linear constraint inherent in SE${}_i$,
optimization problem P$^5_\alpha$ was also solved using the two-step approach,
with the first optimization step determining the class weights, $w_i^{(p)}$.
If $w_i^{(p)} = 0$ for some class $i$, then $R_i^{(p)} = 0$ and SE${}_i = - w_i^{(b)} R_i^{(b)}$, independent of portfolio $p$.
Setting a constraint on SE${}_i$ in such a case makes no sense.
Thus a necessary, though certainly not sufficient, requirement to establish an SE constraint on class $i$ is that $w_i^{(p)}$ be non-negative. 

\begin{figure}[h]
\begin{center}
    \begin{subfigure}[b]{0.6\textwidth} 
    	\includegraphics[width=\textwidth]{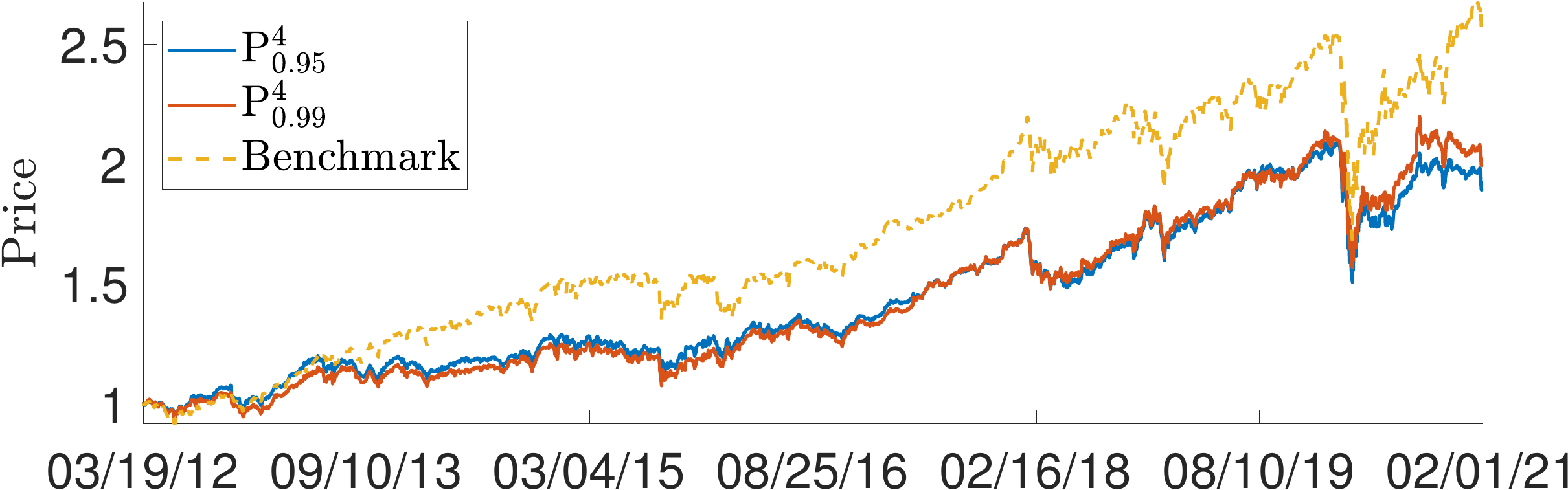}
    \end{subfigure}
     \hfill%
    \begin{subfigure}[b]{0.35\textwidth} 
    	\includegraphics[width=\textwidth]{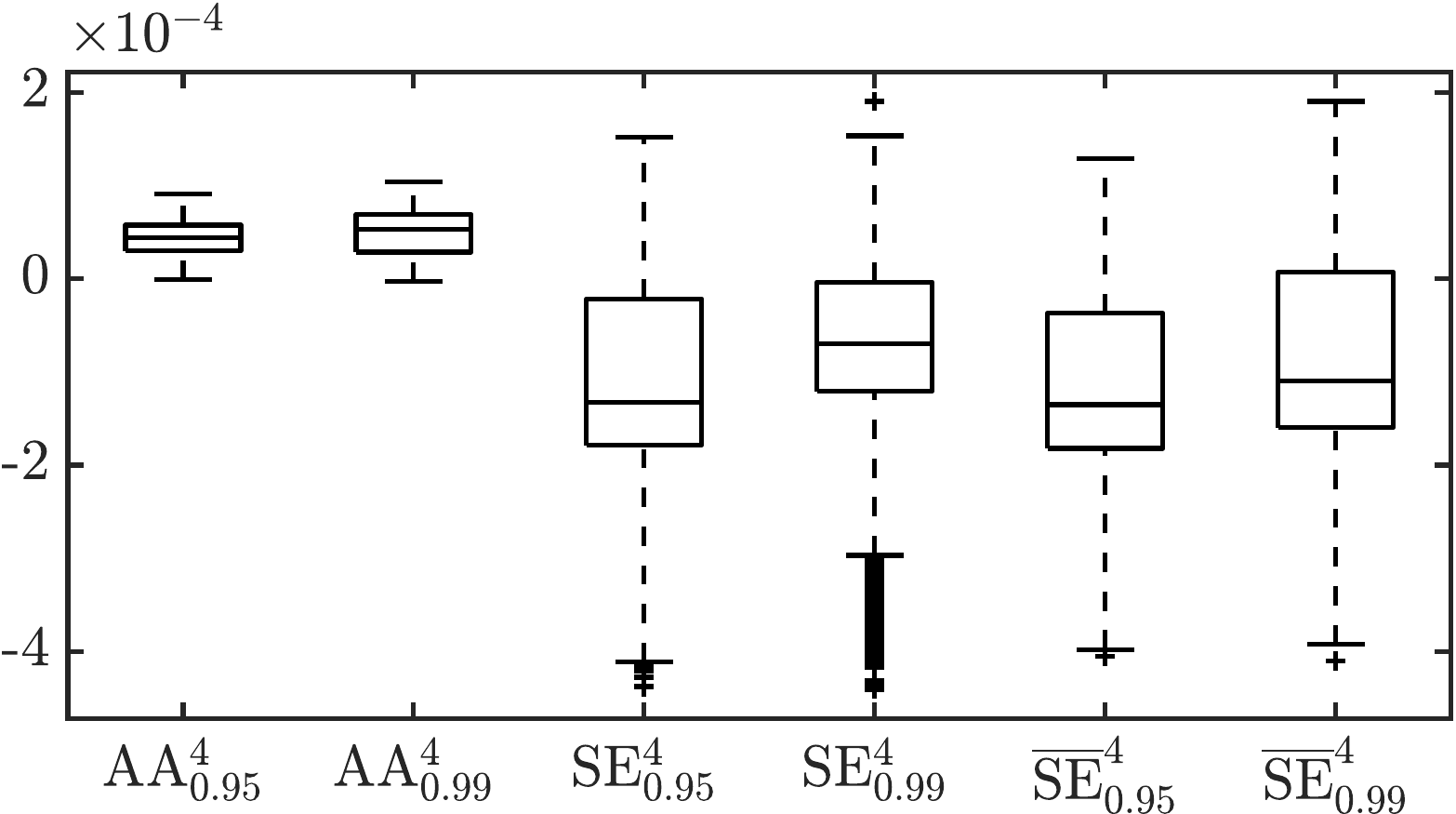}
    \end{subfigure}
    \begin{subfigure}[b]{0.6\textwidth} 
    	\includegraphics[width=\textwidth]{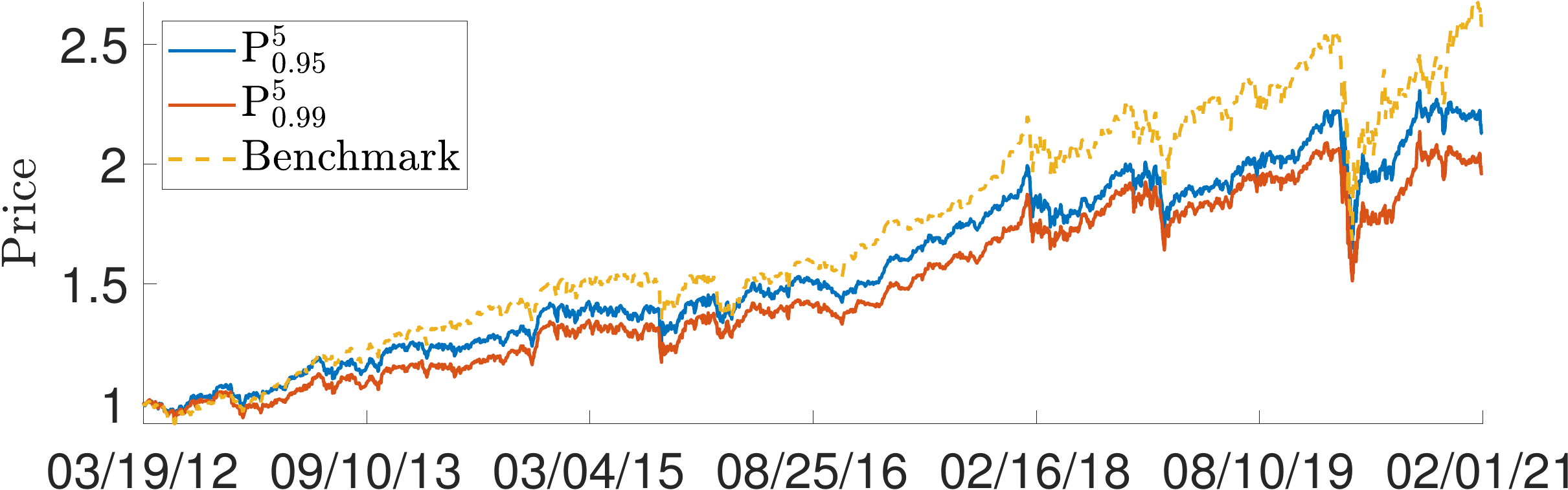}
    \end{subfigure}
     \hfill%
     \begin{subfigure}[b]{0.35\textwidth} 
    	\includegraphics[width=\textwidth]{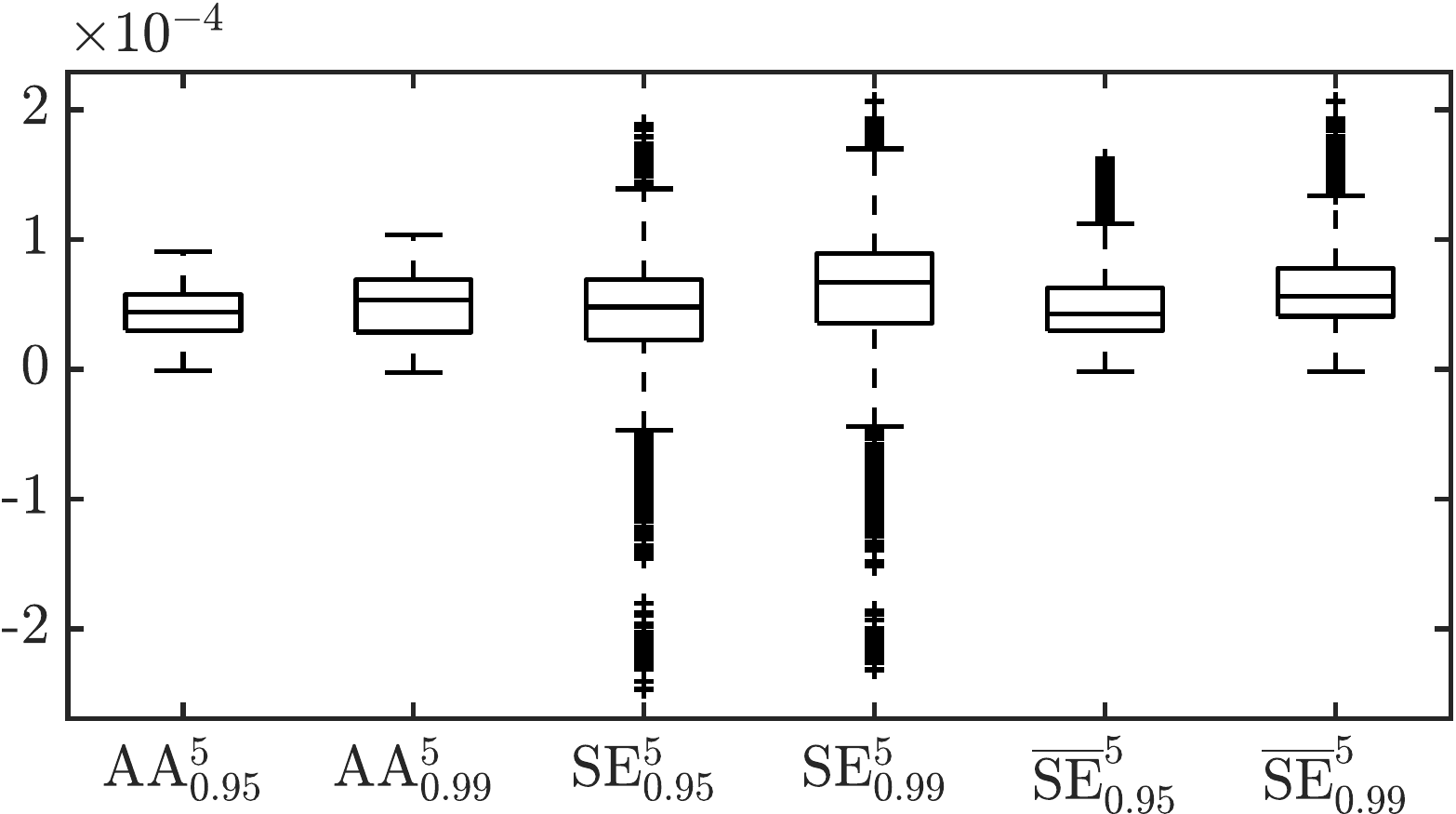}
    \end{subfigure}
    \begin{subfigure}[b]{0.6\textwidth} 
    	\includegraphics[width=\textwidth]{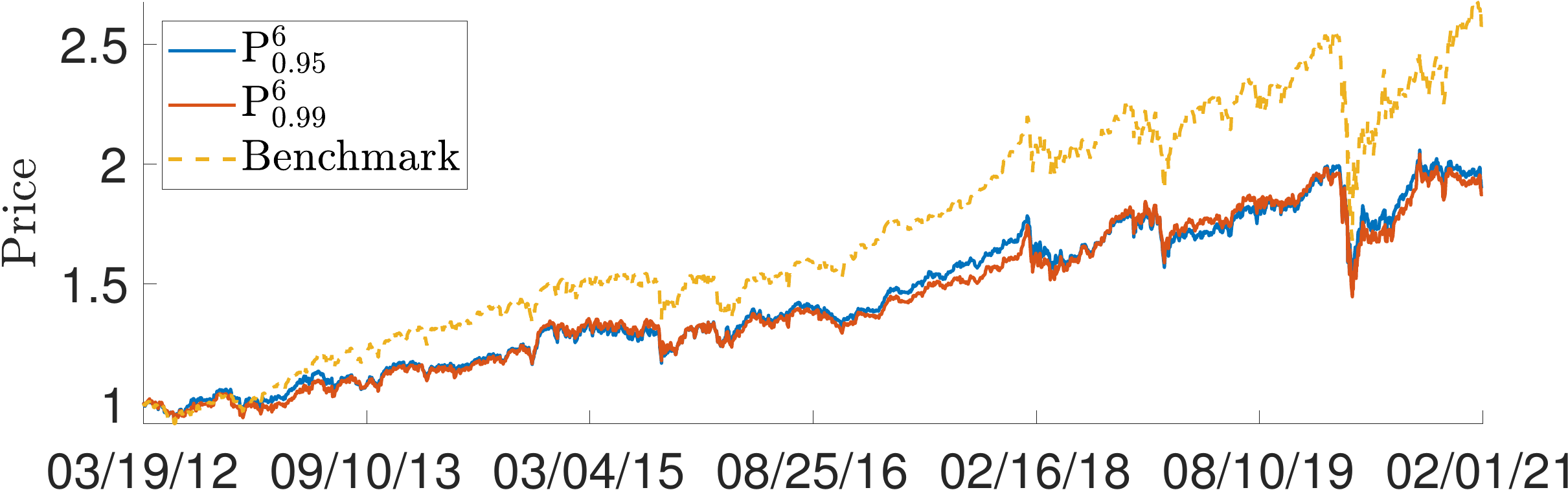}
    	\caption{}
    	\label{fig_P456_price}
    \end{subfigure}
     \hfill%
    \begin{subfigure}[b]{0.35\textwidth} 
    	\includegraphics[width=\textwidth]{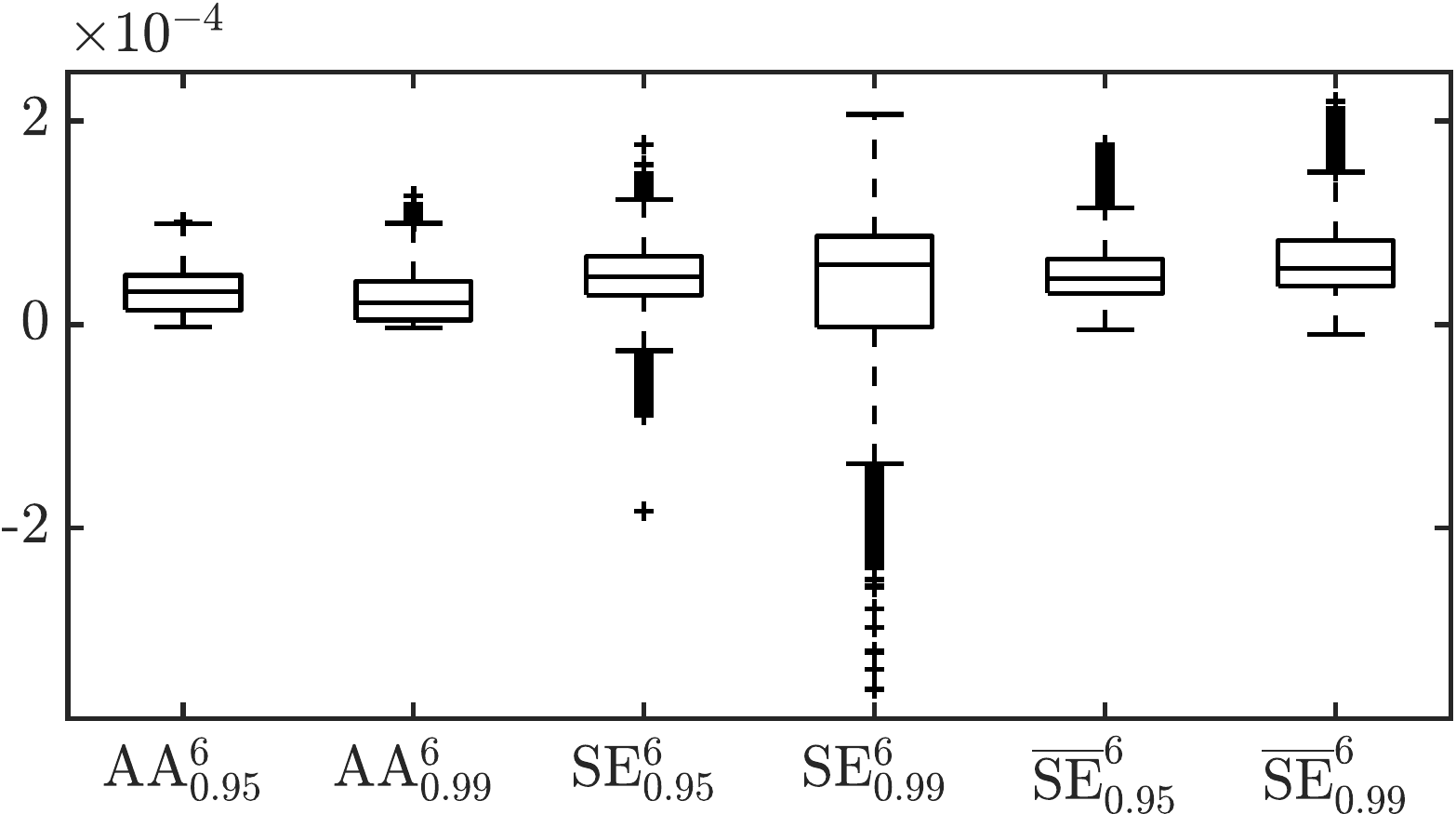}
    			\caption{} 
		\label{fig_P456_box}
    \end{subfigure}
    \caption{(a) Cumulative price resulting from a \$1 investment and
		(b) box-whisker plots of the observed distributions of AA, SE and $\overline{\textrm{SE}}$ values
		for the optimized portfolios P${}_\alpha^4$, P${}_\alpha^5$ and P${}_\alpha^6$.}
    \label{fig_P456}
\end{center}
\end{figure}

Fig. \ref{fig_P456} shows the cumulative price performance and the statistics on the observed distributions
of the performance attributes for these portfolios.
The price performance of P$^4_\alpha$ improved relative to that of P$^1_\alpha$,
with the largest improvement occurring for P$^4_{0.95}$.
Compared to P$^2_\alpha$, price performance improved for P$^5_{0.95}$, while it degraded for P$^5_{0.99}$.
Significant improvements are seen in P$^6_\alpha$ relative to P$^3_{\alpha}$.
The spread of AA$^4_\alpha$ values increased compared to AA$^1_\alpha$.
The spread of AA$^5_\alpha$, SE$^5_\alpha$ and $ \overline{\textrm{SE}}^5_\alpha$ values increased and were more
significantly positive than for P$^2_\alpha$.
Similar observations hold for performance attribution distribution statistics for P$^6_\alpha$.

The risk measures for these portfolios are given in Fig. \ref{fig_RM}.
Compared to their counterparts in optimizations P$^1_\alpha$ through P$^3_\alpha$:
1) with the exception of P$^4_{0.95}$, some improvement in MDD was observed;
2) Sharpe ratios improved, with strong improvements of P$^6_\alpha$ compared to P$^3_\alpha$; and
3) Rachev ratios improve, with the exception of P$^5_\alpha$ compared to P$^2_\alpha$,

To highlight the aggressiveness of optimizations  P$^4_{\alpha}$, P$^5_{\alpha}$  and P$^6_{\alpha}$, Fig. \ref{fig_FR} summarizes,
for each optimized portfolio,
the percentage of the time that no weight solution could be found in the feasibility region.\footnote{
	If no feasibility region solution was found for day $t$,
	we employed a momentum strategy and set $w_{ij}(t) = w_{ij}(t-1)$.\label{fn_no_w}}
``No-solution'' rates are unacceptably high for P$^4_{\alpha}$ through P$^6_{\alpha}$,
indicating that such aggressive constraints should be implemented as soft, rather than hard, constraints.

\begin{figure}[h]
\begin{center}
    \begin{subfigure}[b]{0.6\textwidth} 
    	\includegraphics[width=\textwidth]{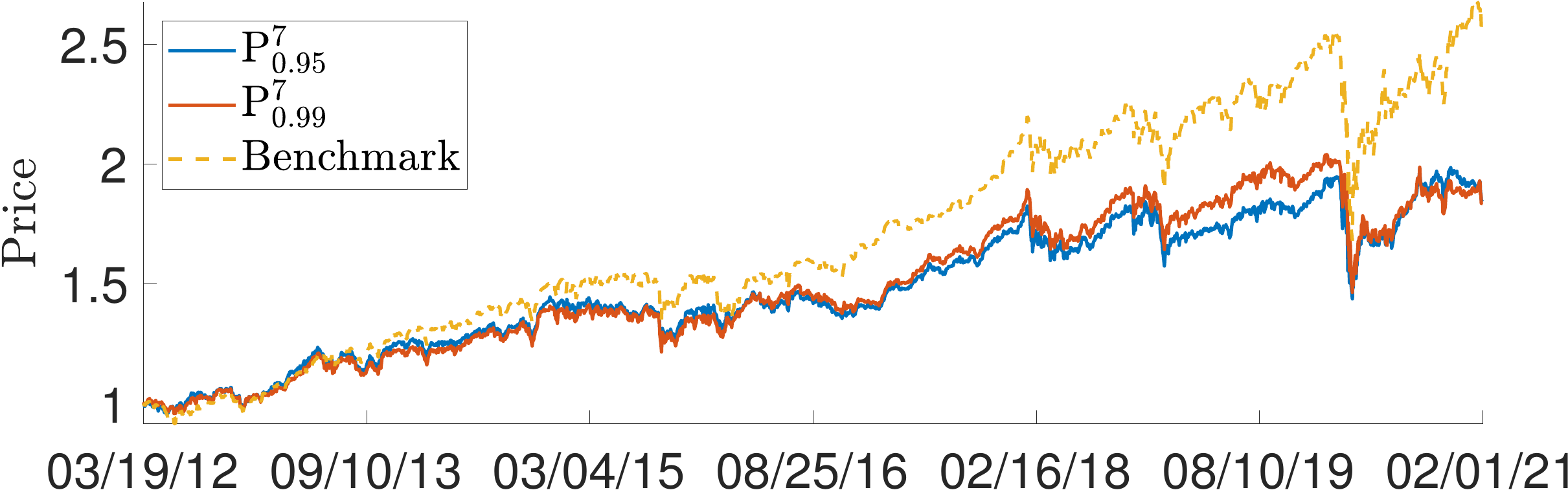}
    \end{subfigure}
     \hfill%
     \begin{subfigure}[b]{0.35\textwidth} 
    	\includegraphics[width=\textwidth]{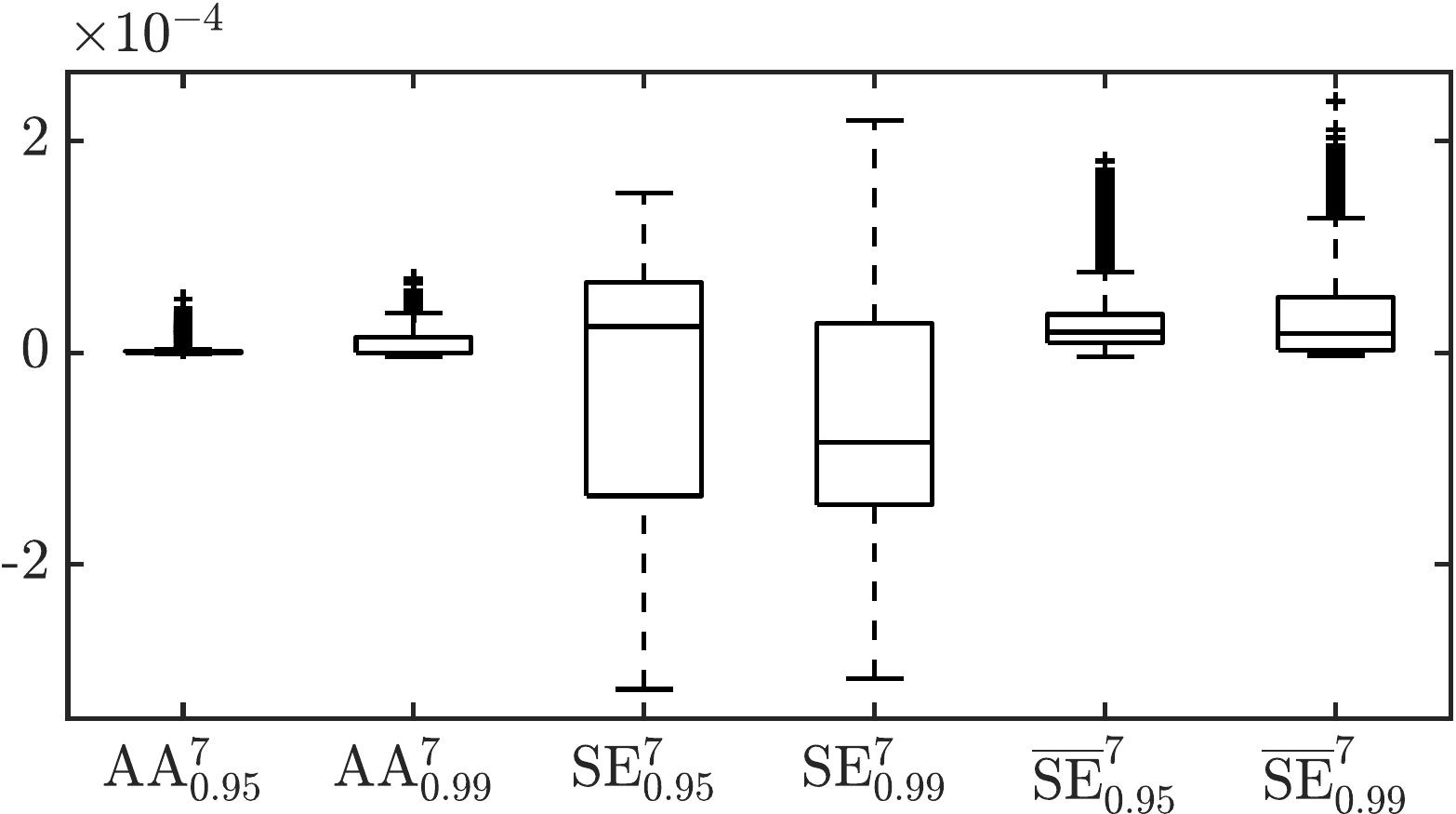}
    \end{subfigure}
    \begin{subfigure}[b]{0.6\textwidth} 
    	\includegraphics[width=\textwidth]{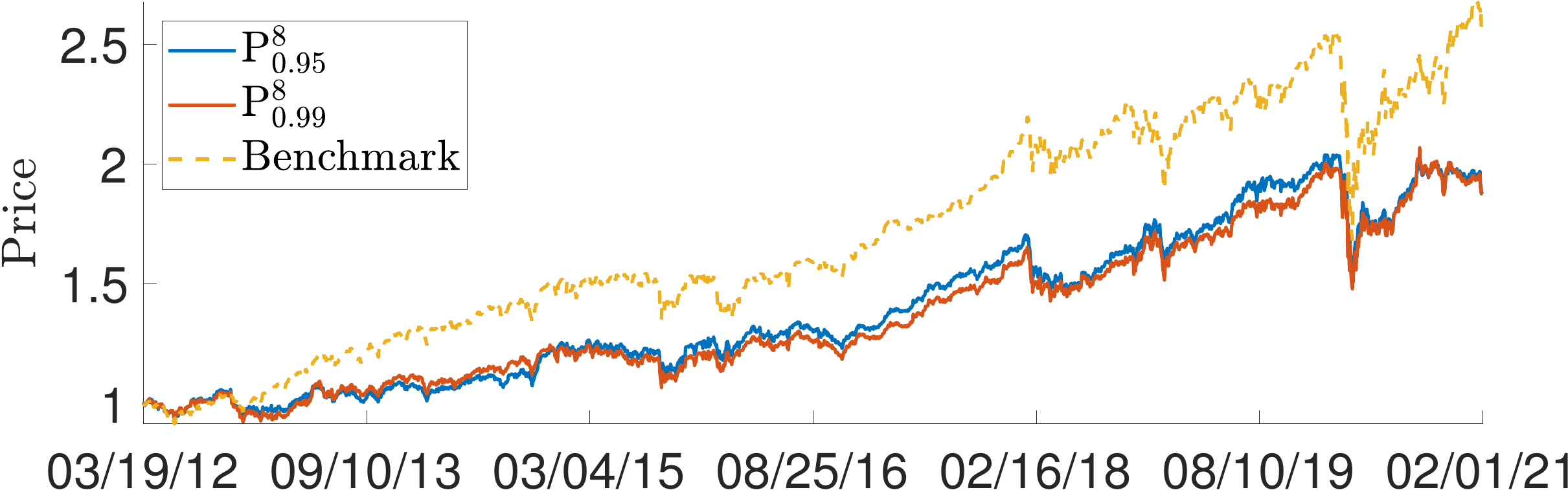}
    	\caption{}
    	\label{fig_P78_price}
    \end{subfigure}
     \hfill%
    \begin{subfigure}[b]{0.39\textwidth} 
    	\includegraphics[width=\textwidth]{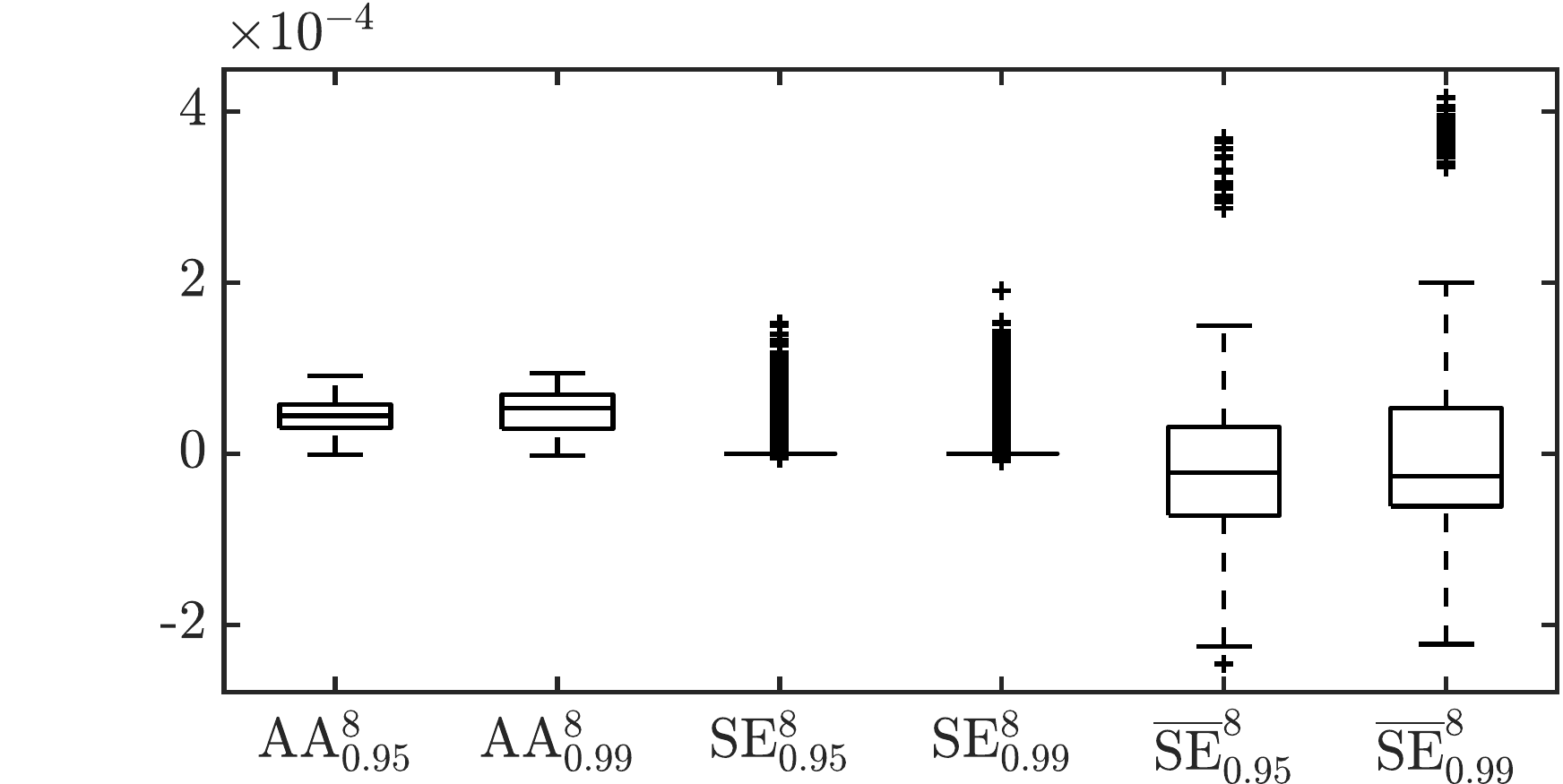}
    	\caption{} 
	\label{fig_P78_box}
    \end{subfigure}
    \caption{(a) Cumulative price resulting from a \$1 investment and
		 (b) box-whisker plots of the observed distributions of AA, SE and $\overline{\textrm{SE}}$ values
		for the optimized portfolios P${}_\alpha^7$ and P${}_\alpha^8$.}
    \label{fig_P78}
\end{center}
\end{figure}
To test the constrictions defining the feasilibity region, we considered two further optimization approaches which mix total and
individual class constraints:
\begin{itemize}[leftmargin=20pt]
\item[P$^7_{\alpha}$:]
(a) $w_{ij}^{(p)}\geq 0,\ \sum_{i,j}w_{ij}^{(p)}= 1$; (b) $0\leq \textrm{AA}$;
	and (c) $0\leq \textrm{SE}_i \textrm{ when } w_i^{(p)} > 0,\ i = 1, \dots, M $.
\item[P$^8_{\alpha}$:]
(a) $w_{ij}^{(p)}\geq 0,\ \sum_{i,j}w_{ij}^{(p)}= 1$; (b) $0\leq \textrm{AA}_i$; and (c) $0\leq \textrm{SE}$.
\end{itemize}
Cumulative price plots for these optimizations are shown in Fig.~\ref{fig_P78};
risk measures are given in Fig.~\ref{fig_RM};
and no-solution rates in Fig.~\ref{fig_FR}.
The price performances of P$^7_{0.95}$ and P$^7_{0.99}$ are very similar and are generally not as good as P$^2_\alpha$ and P$^5_\alpha$.
The price performances of P$^8_{0.95}$ and P$^8_{0.99}$ are also very similar and are generally lag that of P$^7_\alpha$ until 2021.
The wide distribution of negative SE values for P$^7_\alpha$ requires some explanation.
For $p = \textrm{P}^7_\alpha$, $w_1^{(p)} = w_5^{(p)} = 0$ for asset classes 1 and 5
over the period 03/19/12 through 03/04/15.
Thus constraint (c) is not imposed on these classes during this time period.
As negative values for SE$_i$ are, generally, much greater in magnituge than positive SE$_i$ values,
non-constrained time steps skew total values of SE to comparatively large negative numbers.
For the same value of $\alpha$, no-solution rates for P$^7_\alpha$ are comparable to P$^2_\alpha$,
while those for P$^8_\alpha$ are comparable to P$^5_\alpha$.
Of the optimizations P$^2_\alpha$, P$^5_\alpha$, P$^7_\alpha$ and P$^8_\alpha$, Rachev ratios are the lowest for P$^7_\alpha$,
while Sharpe ratios are generally comparable.

\section{Conclusions}
\label{sec4}
\noindent
It is well known that no single optimization method can achieve all goals.
Compared with the ETL optimizations P$^0_\alpha$ which impose no performance attribution constraints,
the price and risk measure performances of the eight performance-constrained ETL$_\alpha$ optimizations,
each computed at two values of $\alpha = 095, 0.99$,
considered in this study lead to the following observations.
\begin{itemize}[leftmargin=10pt]
\item The eight optimizations P$^4_\alpha$, P$^5_\alpha$, P$^6_\alpha$, and P$^8_\alpha$, $\alpha = 095, 0.99$
that seek to simultaneously constrain AA$_i, i = 1, \dots, M$ have the largest rates for which no optimizing solution
is found in the feasibility region.${}^{\ref{fn_no_w}}$
For each risk measure value (MDD, Sharpe ratio, Rachev ratio) five or more of these eight optimizations performed
in the top 50\%.
\item The optimizations P$^1_\alpha$, and P$^3_\alpha$, $\alpha = 095, 0.99$ which had no-solution rates below 0.2\%,
generally registered (at least three out of the four) in the bottom half of the performers for each risk-measure.
\item Of the remaining optimizations, P$^2_\alpha$, and P$^7_\alpha$, which had no-solution rates between 2.5\% and 5.5\%, 
P$^2_{0.95}$ and P$^2_{0.99}$ were in the top half performers in terms of Rachev and Sharpe ratios, while 
P$^2_{0.95}$ and P$^7_{0.95}$ were in the top half performers in terms of MDD.
\item That fact that most of the optimized portfolios outperformed the Rachev ratio of the benchmark while
none of them surpassed the Sharpe ratio of the benchmark may reflect the use of the portfolio standard deviation
in the denominator of the Sharpe ratio which penalizes improved positive returns as strongly as it penalizes negative
returns.
\end{itemize}
These observations, combined with the strong price performance of P$^2_{0.95}$ and P$^2_{0.99}$,
suggest the following conclusions from this study.
\begin{enumerate}
\item There is support for considering optimization using constraints on the total values of AA and SE;
there is less support, in term of both price and risk measure performance, for considering constraints on $\overline{\textrm{SE}}$.
\item Constraining SE and AA produces a larger price performance effect than constraining AA alone.
\item Replacement of the strong constraints $a_{1,i} \le \textrm{AA}_i \le b_{1,i}$ or $a_{2,i} \le \textrm{SE}_i \le b_{2,i},  i = 1, \dots, M$
in optimizations P$^4_\alpha$, P$^5_\alpha$,P$^6_\alpha$, and P$^8_\alpha$  by soft constraints,
or by imposing strong AA$_i$ or SE$_i$ constraints only on a subset of the asset classes,
might drastically reduce no-solution rates while retaining the high risk measure performance of these constrained optimizations.
\item The performance of optimizations P$^2_\alpha$, P$^5_\alpha$, P$^7_\alpha$, and P$^8_\alpha$ should be investigated
further using a non-linear optimization solver to replace the two-step optimization used in this study.
\end{enumerate}

\newpage
\appendix
\renewcommand\thefigure{\thesection.\arabic{figure}}
\setcounter{figure}{0}
\renewcommand\thetable{\thesection.\arabic{table}}
\setcounter{table}{0}
\section{Appendix}

\begin{table}[h]
	\small
	\begin{center}
		\caption{The 30 companies comprising the DJIA${}^a$}
		\label{tab1_DJIA}
		\begin{tabular}{l l l l llll}
			 Ticker &Company&Inception&Weight & Ticker &Company&Inception&Weight\\
			 \omit & \omit & Date & (\%) & \omit & \omit & Date & (\%)\\
			\midrule
			 UNH & UnitedHealth & 10/16/1984 & 7.27  & TRV & Travelers Cos. & 11/16/1975 & 3.01\\
			 GS & Goldman Sachs & 05/03/1999 & 5.98 & NKE & NIKE & 12/01/1980 & 2.96\\
			 HD & Home Depot & 09/21/1981 & 5.88 &APPL & Apple & 12/11/1980 & 2.92\\
			 AMGN & Amgen & 07/16/1983 & 5.23 &JPM & JPMorgan Chase & 03/16/1990 & 2.82\\
			 MSFT & Microsoft & 03/12/1986 & 5.21 & PG & Procter \& Gamble & 01/01/1962 & 2.81\\
			 CRM & salesforce.com & 07/22/2004 & 4.97 & IBM & Int'l Business Mach. & 01/01/1962 & 2.63\\
			 MCD & McDonald's & 07/04/1966 & 4.52 &AXP & American Express & 03/31/1972 & 2.55\\
			 V & Visa & 03/18/2008 & 4.31 &CVX & Chevron & 01/01/1962 & 1.88\\
			 BA & Boeing & 01/01/1962 & 4.26 & MRK & Merck \& Co. & 01/01/1970 & 1.68\\
			 HON & Honeywell Int'l. & 01/01/1970 & 4.25 & INTC & Intel & 03/16/1980 & 1.23\\
			 CAT & Caterpillar & 01/01/1962 & 4.02 &VZ & Verizon Commun. & 11/20/1983 & 1.18\\
			 MMM & 3M & 01/01/1970 & 3.8 & DOW & Dow & 03/20/2019 & 1.15\\
			 DIS & Walt Disney & 01/01/1962 & 3.72 &WBA & Walgreens Boots All. & 03/16/1980 & 1.06\\
			 JNJ & Johnson \& Johnson & 01/01/1962 & 3.54 & KO & Coca-Cola & 01/01/1962 & 1.06\\
			 WMT & Walmart & 08/24/1972 & 3.03 &CSCO & Cisco Systems & 02/15/1990 & 0.99\\
			\bottomrule
			\multicolumn{8}{l}{${}^a$ Bloomberg Professional Services, as of 02/01/2021, 19:57 EST.}\\
		\end{tabular}
	\end{center}
\end{table}

\begin{figure}[h]
    \begin{subfigure}[b]{0.5\textwidth} 
    	\includegraphics[width=\textwidth]{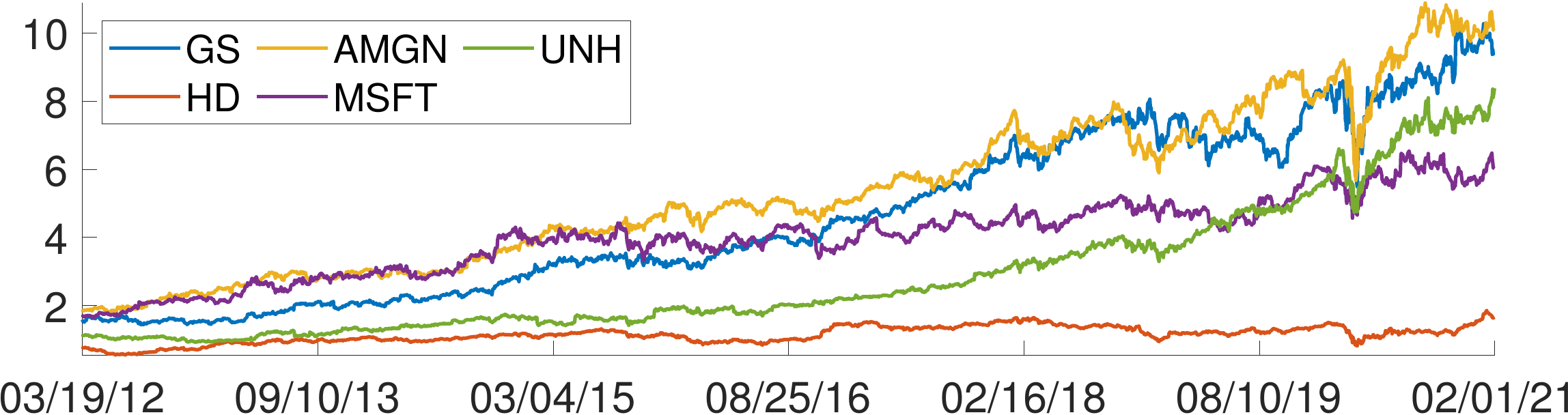}
    	 \vspace{-.66cm}
    \end{subfigure}
    \begin{subfigure}[b]{0.5\textwidth} 
    	\includegraphics[width=\textwidth]{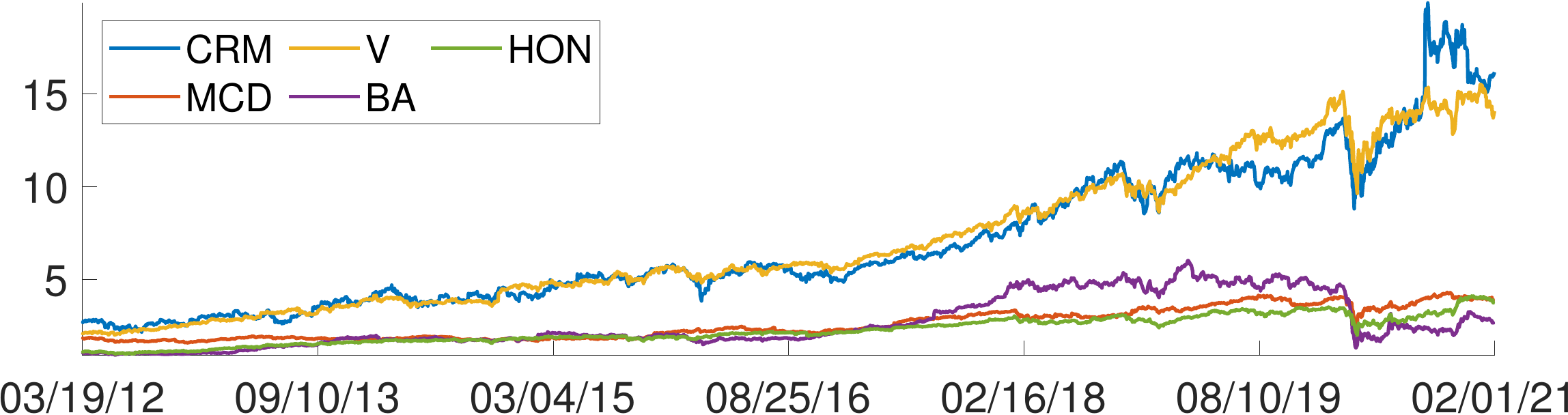}
    	\vspace{-.66cm}
    \end{subfigure}
    \begin{subfigure}[b]{0.5\textwidth} 
    	\includegraphics[width=\textwidth]{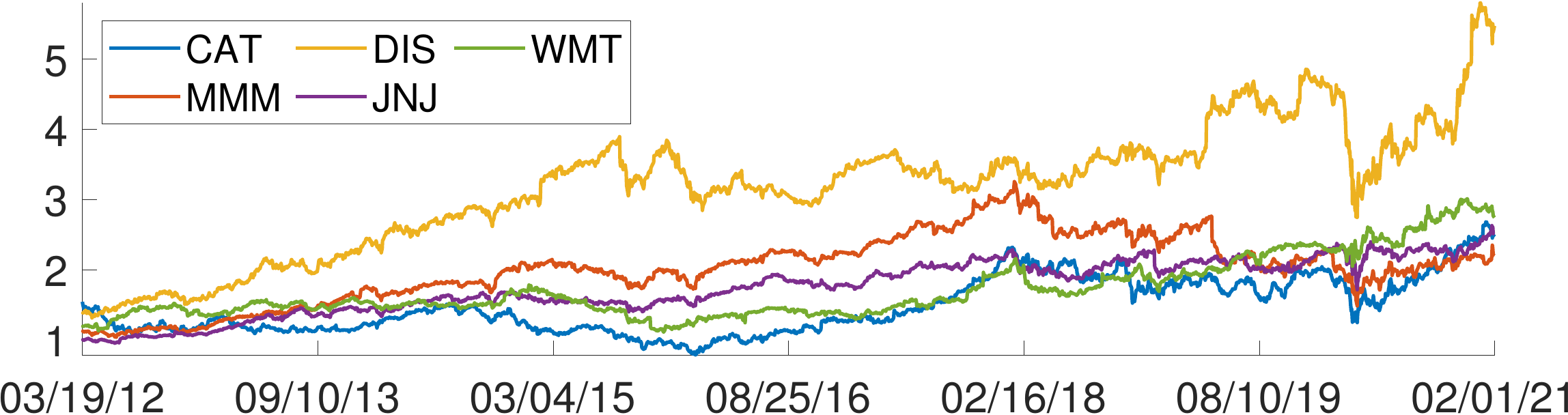}
    	\vspace{-.66cm}
    \end{subfigure}
    \begin{subfigure}[b]{0.5\textwidth} 
    	\includegraphics[width=\textwidth]{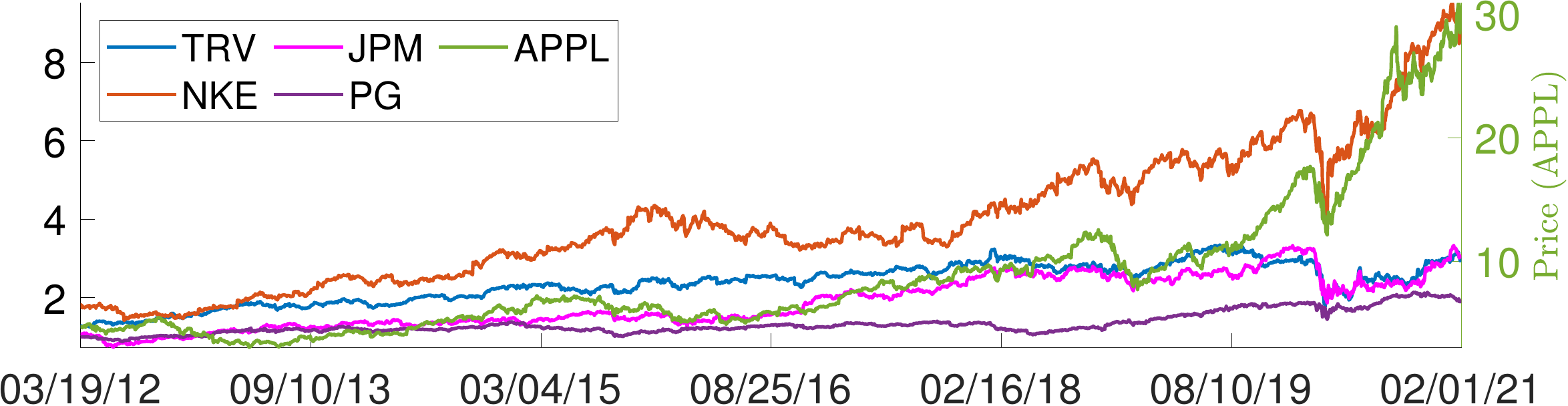}
    	\vspace{-.66cm}
    \end{subfigure}
    \begin{subfigure}[b]{0.5\textwidth} 
    	\includegraphics[width=\textwidth]{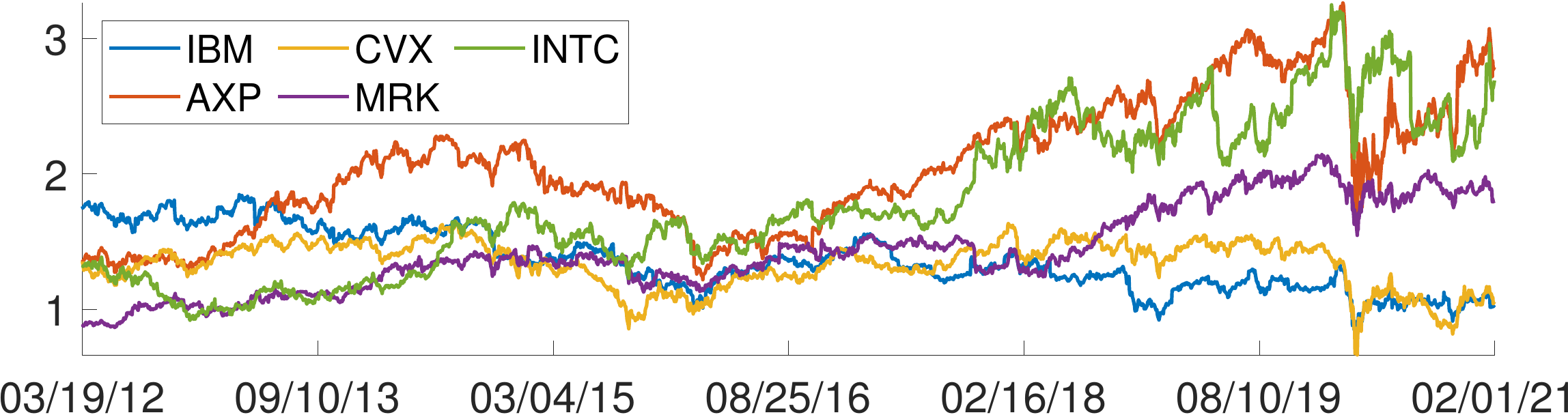}
    \end{subfigure}
    \begin{subfigure}[b]{0.5\textwidth} 
    	\includegraphics[width=\textwidth]{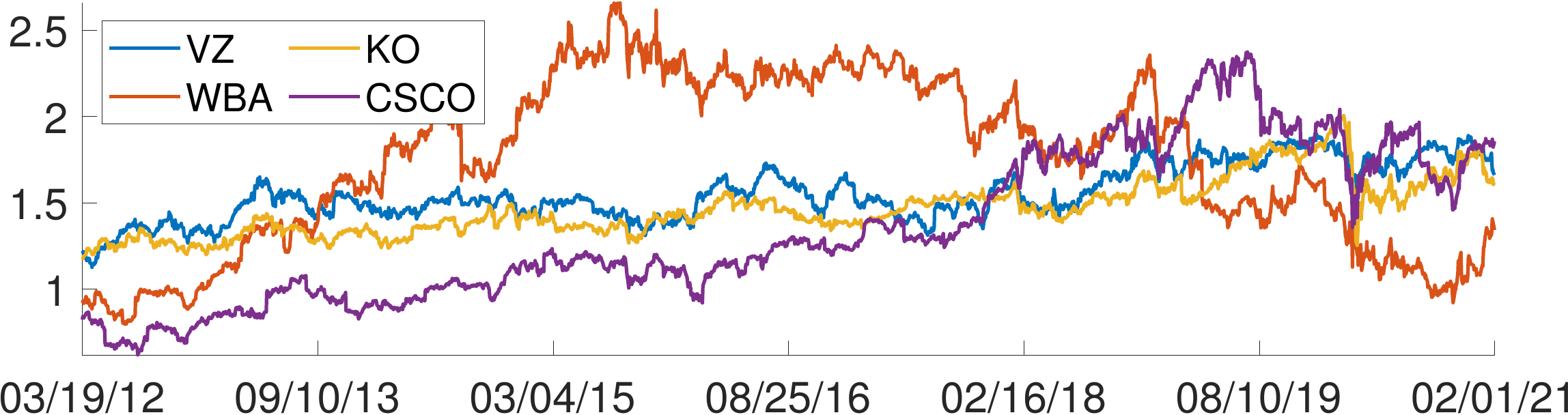}
    \end{subfigure}
    \caption{Cumulative price performance of the stocks in classes (top left) 1 to (bottom right) 6, assuming a \$1 investment on 03/19/2012.}
    \label{fig_Sij}
\end{figure}

\end{document}